\documentclass[10pt,journal]{IEEEtran}
\usepackage{csquotes}
\usepackage{textgreek}
\usepackage{tabularx}
\usepackage{multirow}
\usepackage{multicol}
\usepackage{soul}
\usepackage{graphicx} 
\usepackage{caption}
\usepackage{dblfloatfix}    
\usepackage{textcomp}

\usepackage{booktabs} 
\graphicspath{{figs/}}
\usepackage{braket}
\usepackage{subcaption}
\usepackage{cite}
\usepackage{float}
\usepackage{amsmath,amssymb,amsfonts}
\usepackage{algorithmic}
\usepackage{xcolor}
\def\BibTeX{{\rm B\kern-.05em{\sc i\kern-.025em b}\kern-.08em
    T\kern-.1667em\lower.7ex\hbox{E}\kern-.125emX}}

\definecolor{OliveGreen}{cmyk}{0.64,0,0.95,0.40}

\begin{document}


\title{Quantum PUF for Security and Trust in \\ Quantum Computing}

\author{Koustubh Phalak, Penn State,
        Abdullah Ash- Saki, \IEEEmembership{Student Member, IEEE,} Penn State,
        Mahabubul Alam, \IEEEmembership{Student Member, IEEE,} Penn State,
        Rasit Onur Topaloglu \IEEEmembership{Senior Member, IEEE,} IBM,
        and Swaroop Ghosh, \IEEEmembership{Senior Member, IEEE}, Penn State}

\maketitle
\begin{abstract}
Quantum computing is a promising paradigm to solve computationally intractable problems. Various companies such as, IBM, Rigetti and D-Wave offer quantum computers using a cloud-based platform that possess several interesting features namely, (i) quantum hardware with various number of qubits and coupling maps exist at the cloud end that offer different computing capabilities; (ii) multiple hardware with identical coupling maps exist in the suite; (iii) coupling map of larger hardware with more number of qubits can fit the coupling map of many smaller hardware; (iv) the quality of each of the hardware is distinct; (v) user cannot validate the origination of the result obtained from a quantum hardware. In other words, the user relies on the scheduler of the cloud provider to allocate the requested hardware; (vi) the queue of quantum programs at the cloud end is typically long and maximizing the throughput, which is the key to reducing costs and helping the scientific community in their explorations. The above factors motivate a new threat model with following possibilities: (a) in future, less-trustworthy quantum computers from 3rd parties can allocate poor quality hardware to save on cost or towards satisfying their falsely-advertised qubit or quantum hardware specifications; (b) the workload scheduling algorithm could have a bug or malicious code segment which will try to maximize throughput at the cost of allocation to poor fidelity hardware. Such bugs are possible for trustworthy providers; (c) a rogue employee in trusted cloud vendor could try to sabotage the vendor's reputation by degrading the user compute fidelity just by tampering with the scheduling algorithm or rerouting the program; (d) a rogue employee can steal information by redirecting the programs to a 3rd party quantum hardware where they have full control. If the allocated hardware is inferior in quality, the user will suffer from poor quality result or longer convergence time. 
We propose two flavors of a Quantum Physically Unclonable Function (QuPUF) to address this issue- one based on superposition and another based on decoherence. Our experiments on real quantum hardware reveal that temporal variations in qubit quality can degrade the quality of the proposed QuPUF. We add a parametric rotation to the QuPUF for stability. Experiments on real IBM quantum hardware show that the proposed QuPUF can achieve inter-die Hamming Distance (HD) of 55\% and intra-HD as low as 4\%, as compared to ideal cases of 50\% and 0\% respectively. The proposed QuPUFs can also be used as a standalone solution for any other application.

\end{abstract}

\begin{IEEEkeywords}
Quantum Computing, Security, Quantum PUF.
\end{IEEEkeywords}

\pagestyle{plain}
\section{Introduction}
\IEEEPARstart{Q}{uantum} computing can solve computationally intractable problems in domains e.g., finance, traffic flow and power grid by exploiting superposition and entanglement properties. Various qubit technologies are being explored including superconducting, Ion Trap and single electron by academia and industry to develop scalable quantum computers. While the best scalable quantum technology is an active area of research, design community is exploring the quantum computers offered by various companies such as, IBM, Rigetti and D-Wave to solve optimization problems. Currently, the access to quantum computers is provided through cloud-based platform where a suite of quantum computers are available for the users to solve their problems. The users can compile their circuits for a particular hardware and transmit to the cloud which enters a queue. The scheduling algorithm allocates 
\begin{figure}[t]
    \centering
    \includegraphics[width=0.80\columnwidth]{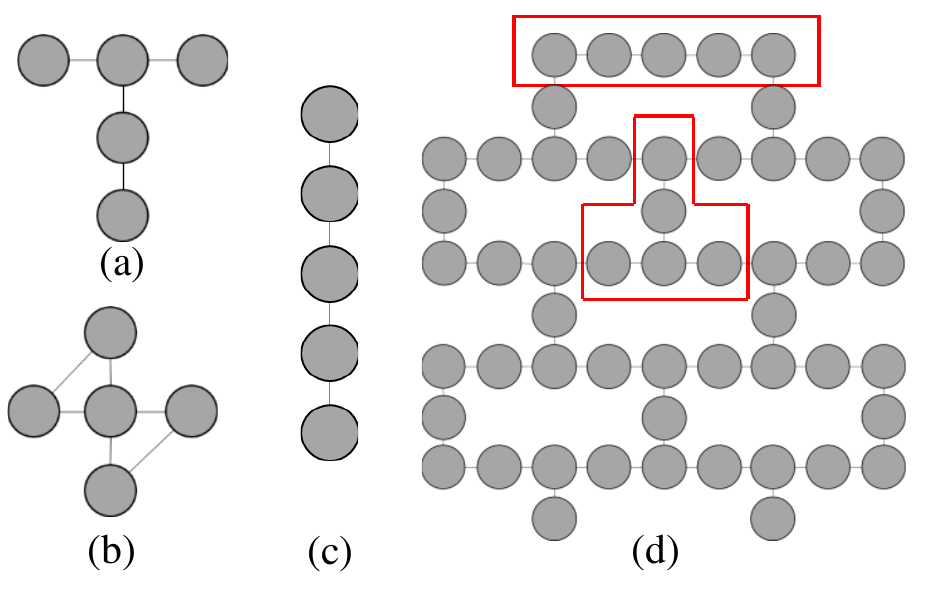}
    \caption{The structures of various quantum computers, (a) ibmq\_london, ibmq\_burlington, ibmq\_essex, ibmq\_vigo, etc.; (b) ibmq\_yorktown; (c) ibmq\_santiago; and, (d) ibmq\_rochester. It can be noted that rochester consists of several isomorphic graphs of ibmq\_london, ibmq\_burlington, ibmq\_santiago etc. Therefore, ibmq\_rochester hardware can accommodate multiple workloads meant for ibmq\_london, ibmq\_santiago etc. in parallel.}
    \label{fig:coupling_map}
\end{figure} 
the programs from the queue using a pre-defined allocation policies such as, fair share allocation ~\cite{das2019case}. Once the experiment is concluded, the results are sent back to the user. Since the noisy computers are less powerful and limited in the number of qubits, various hybrid algorithms such as, Quantum Approximate Optimization Algorithm (QAOA) and Variational Quantum Eigensolver (VQE) are pursued where a classical computer drives the parameters of a quantum algorithm/circuit iteratively. The goal of the classical computer is to find the right set of parameters that can drive the quantum algorithm towards the optimal solution for a given problem. For high-quality hardware with reliable qubits, the algorithm is expected to converge faster i.e., with less number of iterations. However, the quantum computers with more number of qubits and/or high quality qubits typically come at a higher cost. Therefore, securing the hardware with desired quality is very important to solve a certain problem within a desired deadline. However, the user gets little/no visibility about the hardware allocated to the program in the existing setup. This gives rise to following new challenges: \vspace{-1mm}



\textbf{Hardware suite:} The cloud service provider may possess multiple hardware with varied degree of computing capability i.e., number of qubits and coupling map. For example, IBM Quantum has access to multiple quantum hardware like ibmq\_london, ibmq\_burlington, ibmq\_essex, ibmq\_santiago and ibmq\_rochester (Fig. \ref{fig:coupling_map}).

\begin{figure} [H] 
\centering
\vspace{-1em}
    \includegraphics[width=\linewidth]{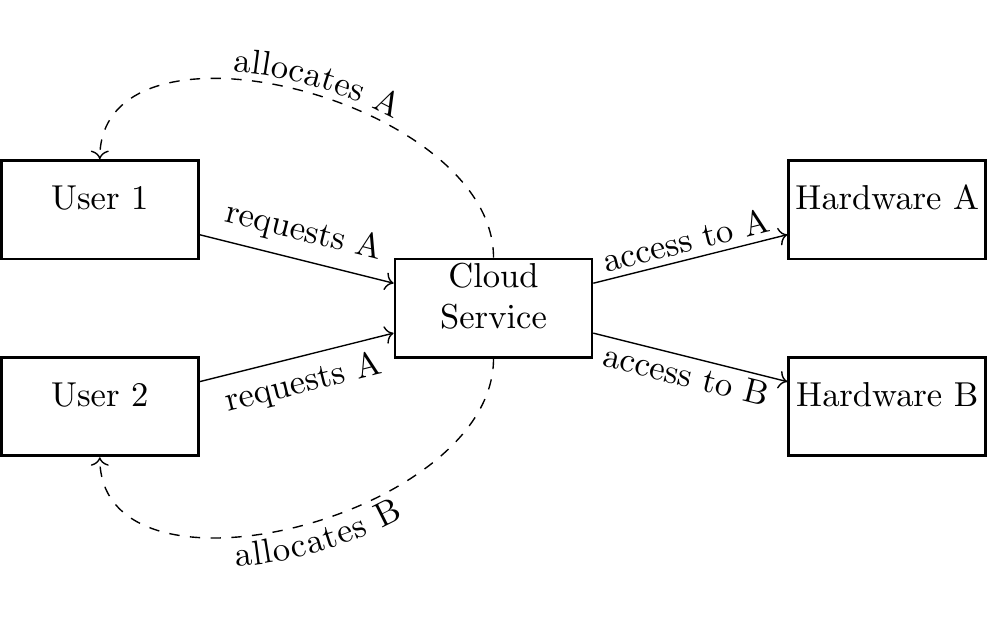}
        
     \caption{Conceptual attack model where both users request for hardware A (with superior quality) but user-2 gets access to hardware B. 
     }    
 \label{fig:attack_model}
 \vspace{-0.5em}
\end{figure}

\textbf{Multiple choices for user-specified coupling map:} The scheduler at the cloud service provider end may have multiple hardware with identical coupling maps in the suite e.g., ibmq\_london, ibmq\_essex, ibmq\_vigo etc. (Fig. \ref{fig:coupling_map}). Structurally, one cannot differentiate them from each other.

\textbf{Isomorphic coupling maps in larger hardware:} The larger quantum hardware with more number of qubits possess multiple isomorphic coupling maps of many smaller hardware e.g., ibmq\_rochester has many T-shaped coupling maps similar to ibmq\_london as specified in one of the boxes in Fig. \ref{fig:coupling_map}.

\textbf{Quality and cost differences:} Each of the hardware is distinct in terms of computation capability, quality of qubits, and cost. In general, the larger qubit hardware is more costly; the cost can depend on the qubit quality for identical hardware size.

\textbf{Allocation/scheduling policy:} The queue of quantum programs at the cloud end is typically long and maximizing the throughput is the key to reduce costs and help the scientific and industrial research communities in their explorations. The hardware scheduling policy typically employs a vendor-selected metric (throughput or first-in-first-out) to allocate the program to the hardware.

\textbf{Validation of the non-repudiation of results:} The user cannot validate the origination of the result obtained from a quantum hardware. In other words, the user trusts the scheduler of the cloud provider to allocate the requested hardware to his workload. 

Furthermore, quantum computers are being developed by multiple entities, some of which may be less trustworthy. In future, cloud-based quantum computing is expected to be offered by both trusted and less-trusted cloud vendors (that are located in less-trusted countries, for example). Performing reliable/trustworthy computing using these cloud-based quantum computers is an important step towards expanding their application space.    

\textbf{Proposed attack model:} Consider the situation in Fig.~\ref{fig:attack_model} where User 1 (\textbf{U1}) and User 2 (\textbf{U2}) can access two quantum hardware \textbf{A} and \textbf{B} through the Cloud Service provider (\textbf{CS}) by paying certain service fee. 
Hardware A is relatively superior to B (in terms of error rates, for example), so both U1 and U2 would like to run their program on A. The motivation is to obtain high quality results as well as to reduce the cost if the problem is solved quickly. 
However, only one user can be assigned hardware A at a time i.e., U1 in this case. The scheduler in the CS has the option to make U2 wait or allocate it to B to maximize the throughput or increase user's cost (due to a malicious code segment or just to reduce the wait queue depth). As a result, U2 will suffer from poor quality/incorrect results due to inferior hardware and may also end up paying more. To address this problem, we propose a Quantum Physically Unclonable Function (QuPUF) program to be sent to the hardware to establish its identity first. The program will be allocated to the quantum computer (either the desired hardware or a different hardware) and the response will be sent back to the user. The user will match the response with the registered responses of the desired hardware. Since each quantum hardware has its own unique characteristics (e.g., single-/two-qubit gate error rates, decoherence and dephasing times), the responses of each hardware will be unique. Therefore, the user will be able to validate the identity of the hardware before sending his actual workload. The proposed QuPUF can be sent prior to the actual workload to establish trust (easy) or it can be embedded in the user workload itself (complex, discussed in Section V). \textit{To the best of our knowledge, this the first effort to establish the identity (trust) of a quantum computer}. Note that although we associated the proposed QuPUFs with an attack model, they can also be used as a standalone security and trust anchors.

\textbf{Paper contributions:} We, (i) propose a new attack model and possible modes of attack; (ii) propose 2 flavors of QuPUFs to counter this attack model, (iii) study the stability of the QuPUF and propose to add/optimize the rotation angle of the QuPUF circuit to enhance the stability; (iv) introduce digitization of the response and optimized the bit-precision for improved inter- and intra-HD of the QuPUFs. 

The paper is organized as follows. In Section II, we describe the background on quantum computing and PUF. The attack model is presented in Section III whereas the implementation details of various QuPUFs are provided in Section IV. The limitations of the proposed QuPUFs are discussed in Section V. Conclusion is drawn in Section VI. 

\section{Background} 
In this section, we discuss relevant background on quantum computing, PUF and quantum PUF.

\subsection{Quantum computing concepts}
{\bf{Qubits:}}
A qubit is the building block of a quantum computer. It stores data as quantum state. Unlike classical bits, qubit can be in a superposition state, i.e., a combination of 0 and 1 at the same time.

{\bf{Quantum gates:}}
Quantum gates are the operations that modulate the state of qubits and thus, perform computations. Quantum gates can work on a single qubit (e.g., X (NOT) gate) or on multiple qubits (e.g., 2-qubit CNOT gates). Physically, they are realized using pulses (e.g., laser pulse in Ion Trap qubits, RF pulse in Superconducting qubit, etc.).

{\bf{Errors in noisy quantum computers:}}
Present quantum computers suffer from various error modes such as, gate error, decoherence, readout error, single qubit error, two qubit error and crosstalk. Due to gate error the logical operation of a gate suffers certain probability of error. Qubits spontaneously interact with the environment and lose states which is known as decoherence. Due to imperfections in readout circuitry, qubits can suffer from bit-flips leading to readout errors. There are errors defined based on the type of gate. The errors caused by single qubit gates (like Hadamard gate for instance) is called single qubit error, and errors caused by two qubit gates (like CNOT gate) is called two qubit error. Finally, parallel gate operations on different qubits can affect each others' performance which is known as crosstalk. The rates of these errors vary among qubits and hardware which can be used as a signature to identify a particular hardware.

\textbf{Various factors causing errors: } 
There could be many reasons behind errors including manufacturing imperfections, control error, thermal gradient, environmental interaction, poor microwave hygiene, etc \cite{krantz2019quantum}. Due to manufacturing imperfections, there can be defects/charge traps, and it leads to charge noise, which is a source of gate error. The control errors may stem from incorrectly calibrated gate pulses which may lead to under- or over-rotation of qubits, leakage to non-computation states, etc. For example, the quantum NOT (X) gate is realized by a $90^{\circ}$ rotation around X-axis. A microwave pulse of certain amplitude, shape, and duration is applied to the qubit to drive this rotation. If the amplitude/shape/duration is incorrectly calibrated, the rotation will be less (under) or more (over) than the intended $90^{\circ}$ leading to gate error. Qubits are ideally 2-level systems. the ground state (0) and $1^{st}$ excited state (1) make up the computational space. In Transmon qubits, there is a certain energy difference between 0 and 1 states usually denoted by a frequency $f_{01}$. A qubit is usually driven by a microwave pulse (gate pulse) of frequency $f_{01}$ which will initiate transition between computational 0 and 1 states. However, in practical qubits like Transmon there are higher energy states like $2^{nd}$ excited state, $3^{rd}$ excited state, etc. beyond these two states.  In case of Transmons, the energy difference between $1^{st}$ excited state and $2^{nd}$ excited state, $f_{12}$, is close to $f_{01}$ (known as low anharmonicity). Due to this closeness of frequencies or low anharmonicity and imperfection in control signal, a qubit intended to be driven by $f_{01}$ may jump out of 0 and 1 computational space and get excited to $2^{nd}$ excited state (known as leakage). Qubits are cooled down to cryogenic temperature and it is expected to have a homogenous temperature across the device. However, due to localized heating, there can be a thermal gradient. This thermal gradient is a source of decoherence \cite{spilla2014measurement}. Qubits are very susceptible to noisy environment like stray magnetic fields, heating, etc. For example, a qubit can absorb energy from environment and get excited to non-target state. Therefore, quantum computers are shielded and operated in very controlled environment. However, the solutions are not perfect yet and there are engineering challenges to prevent environmental effects completely. The microwave signal lines may suffer from photon number fluctuations \cite{krantz2019quantum} which causes stark shift. Due to stark shift, the operating frequency ($f_{01}$) of a qubit change. If the operating frequency of a qubit is different than the frequency it is driven, it gives rise to incorrect operation and gate error.

Errors (excluding crosstalk) are not dependent on the number of parallel operations, but rather on individual gate operations. Number of parallel operations give rise to crosstalk error. However, parallel single qubit gates do not incur significant crosstalk. Even this can be mitigated by serializing the gates e.g., by adding delays in the circuit like idle gates or barriers. Reliability is dependent on thermal variation rather than crosstalk. Due to this, the single qubit gate error changes over time, which gives rise to temporal variation, and contributes towards the intra-HD.

\subsection{Physically Unclonable Function (PUF)}

PUF \cite{pappu2002physical} is a physical object which cannot be cloned. It acts a good security measure since the adversary cannot clone the characteristics of the PUF accurately.
PUFs exploit the characteristics which are unique due to the variation in the manufacturing process. Some examples include SRAM random initialization \cite{guajardo2007fpga}, thin-film resonators \cite{skoric2008randomized}, dielectric properties of security coatings \cite{tuyls2006read}, delays in integrated circuits \cite{gassend2002silicon}, etc. PUFs work on the principle of challenge and response. 
The user can provide a challenge to the PUF and obtain the corresponding response for authentication. The correctness of the response is validated by matching it from a database with registered challenges/response pairs (CRP) for the device. The PUFs can be categorized into based on the number of CRP, namely strong (exponential CRP e.g., arbiter PUF) and weak PUF (linear CRP e.g., SRAM PUF). Two important properties of PUF are, (i) inter-die Hamming Distance (HD) which measures the change in response between all pairs of identical chips for the same challenges\footnote{In quantum context that we introduce next, these would be gate error, decoherence, readout error, and crosstalk}. The ideal value of inter-HD should be 50\%; and, (ii) intra-die HD which measures the change in response with respect to time under temporal variation (due to noise, temperature and voltage fluctuations and aging). The ideal value of inter-HD should be 0\%. 


\subsection{Quantum PUF}
A common drawback of the classical PUFs is that the unique parameter created by the process variation is uncontrollable. As a result, even a slight change in the parameter cannot be reverted back to the original value. This can have an undesirable impact on the response for the challenges. Quantum PUFs have been proposed to address the above challenge by providing some controllable unique parameters.
For instance, the concept of quantum confinement is employed to produce a unique signature by exploiting the fluctuations in quantum tunneling measurements inside resonant tunneling diodes (RTD) \cite{roberts2015using}. A quantum secure authentication (QSA) using illumination with a light pulse and checking the shape of the reflected light is also proposed \cite{goorden2014quantum}. However, both of these techniques heavily depend on quantum physics and are not applicable in the proposed application. 
A quantum challenge and quantum state readout of a classical optical PUF is proposed in \cite{vskoric2012quantum}. Another work presents a quantum PUF which employs quantum properties of quantum device to establish a secure communication channel against quantum cryptographic attacks \cite{arapinis2019quantum}. While some theoretical foundations are discussed,  practical application, method, and circuits that relate to trustworthy computing in a cloud environment are completely unaccounted for.

\subsection{Hardware variability}
In case of quantum computers, hardware variability manifests as variable “hardware errors”, more specifically gate error rates, decoherence times (e.g., T1-relaxation), etc, across different quantum chips. In classical computing domain, two same chips may perform differently e.g., they may have different gate-delays due to manufacturing variations. This means hardware variability is manifested as variable gate-delays in classical chips. Likewise, variability between two quantum computing hardware is demonstrated/accounted as differences in gate error rates, decoherence (T1) times, etc. Therefore, similar to gate delays in classical chips - which can be used as a representation of variability and as a hardware signature – gate error-rates and decoherence times can be used as a representative of variability and hardware signatures in the quantum domain.

For example, the Transmon qubits used in the IBM machines is ultimately a solid-state device. “Trapped charge” in defects is a ubiquitous phenomenon in all solid-state devices alike. This trapped charge is a reason behind gate error in Transmon qubits \cite{krantz2019quantum}. Due to manufacturing variations, the number of defects will vary among qubits leading to variable gate errors which will be exploited to design the PUF. 

\begin{figure}
    \includegraphics[width=1.0\linewidth]{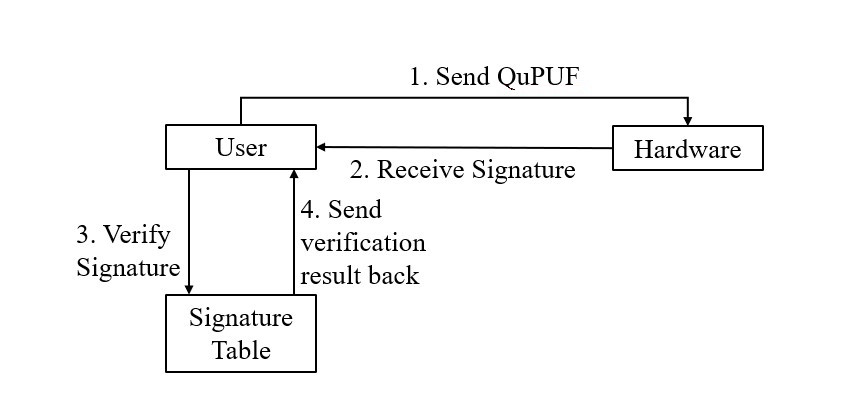}
    \label{fig:qupuf_usage}
    \caption{The usage of QuPUF. Each step has been numbered.}
\end{figure}

\section{Proposed Attack Model} 
In this section, we describe the attack model and various attack scenarios.
\subsection{Basic idea}
In this attack model, the buggy or malicious or 3rd party controlled scheduler is the adversary which fails to allocate the quantum computer requested by the user's program but rather, (i) allocates a different quantum computer with identical coupling map, or (ii) maps the program to a smaller segment of a larger hardware along with other programs running in different segments in parallel. For example, the ibmq\_rochester device in Fig. \ref{fig:coupling_map} (d) contains several T-shaped coupling maps (highlighted in red) where the programs meant for ibmq\_london could be executed. This could be motivated by 
multiple types of scenarios described next. 
For simplicity, we call the quantum computer where the program was supposed to run as `target' quantum computer and the one where the program actually runs as `allocated' quantum computer. 

\subsection{Attack scenarios} \label{obs1}
\textbf{Scenario-1) Throughput maximization}: In this scenario, the scheduling routine attempts to make the best out of the hardware suite to maximize the throughput while adhering to the user-desired coupling map. This could either be intentional (a decision made by the cloud service provider) or due to a bug. If the user specified quantum computer is unavailable due to other prior tasks, the scheduler may divert this job to another quantum computer with the same or greater number of qubits. It is worth noting that if the number of qubits is same, then the computation will depend on the quality of the allocated hardware (which can be poor compared to the target hardware). However, if the number of qubits of the allocated quantum computer is higher than those of the target quantum computer, it is possible that another task is already being run on the allocated quantum computer. Such an arrangement, where two different tasks are being run simultaneously on the same quantum computer will give rise to crosstalk error due to inter-circuit interference \cite{ash2020analysis,murali2020software}. 
The quality of these subgraphs may be worse than the user specified hardware.

\begin{figure*}[!t]
    \begin{subfigure}{0.31\linewidth}
        \includegraphics[width=\linewidth]{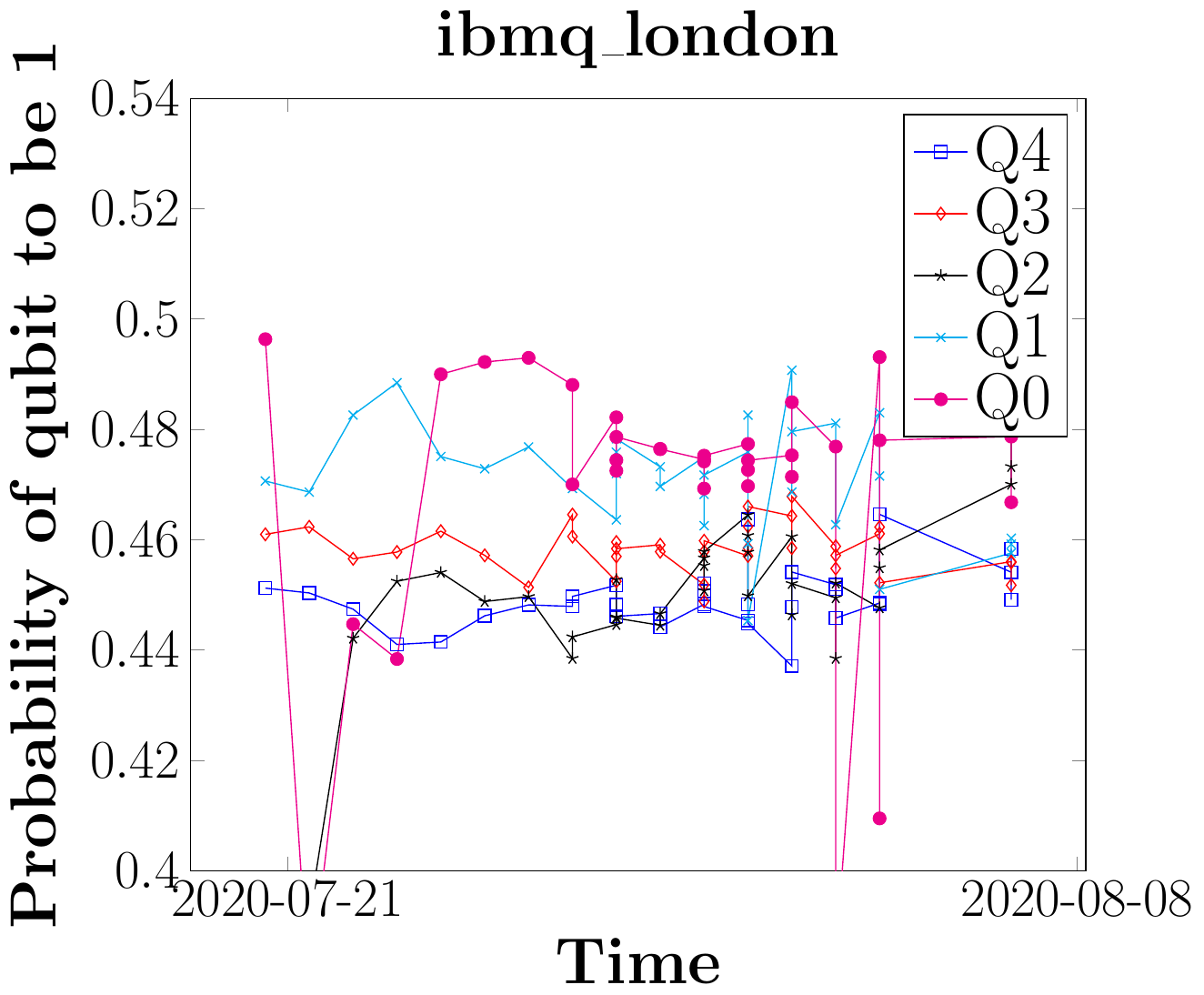}
        \caption{}
    \end{subfigure}
    \begin{subfigure}{0.31\linewidth}
        \includegraphics[width=\linewidth]{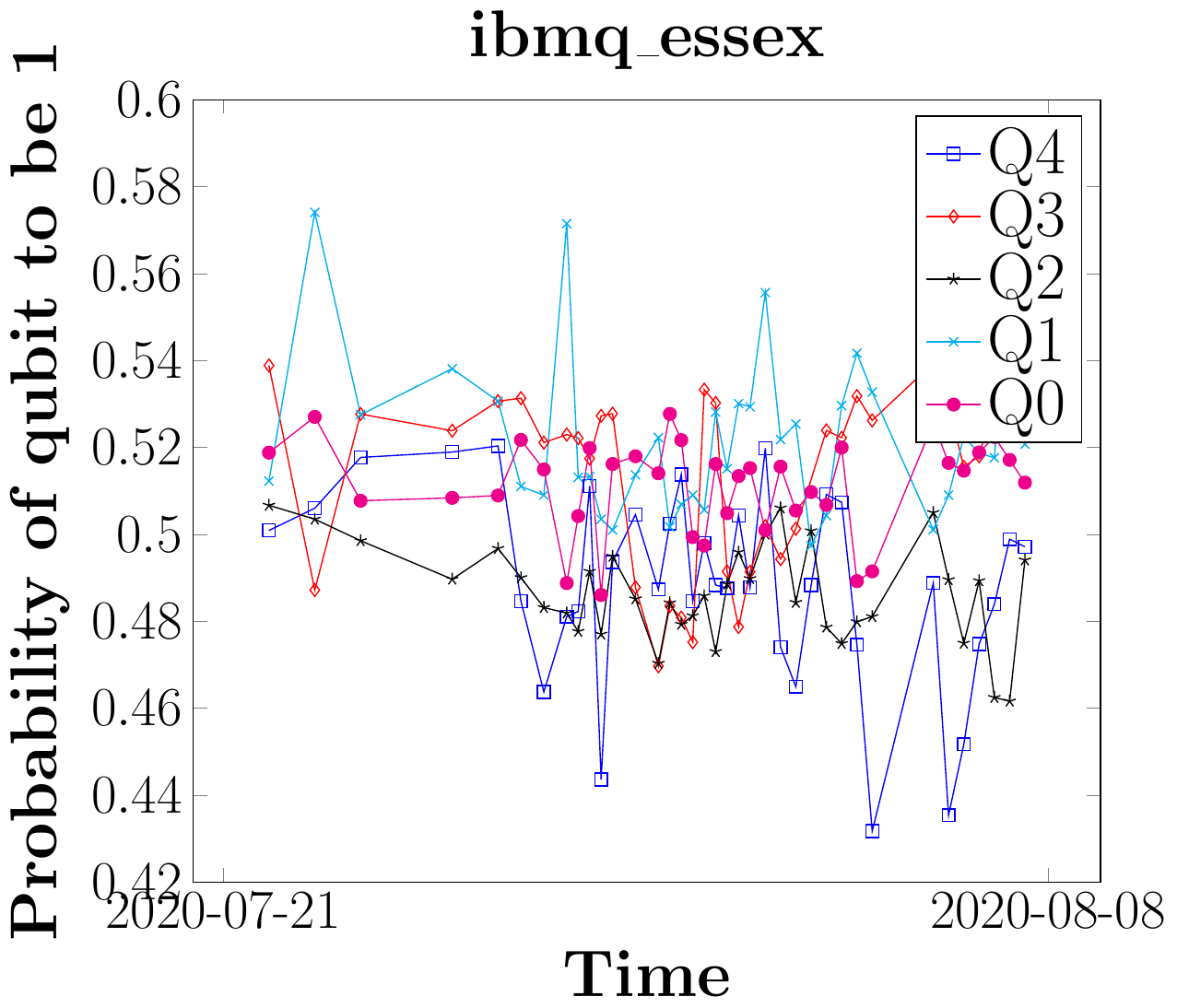}
        \caption{}
    \end{subfigure}
    \begin{subfigure}{0.31\linewidth}
        \includegraphics[width=\linewidth]{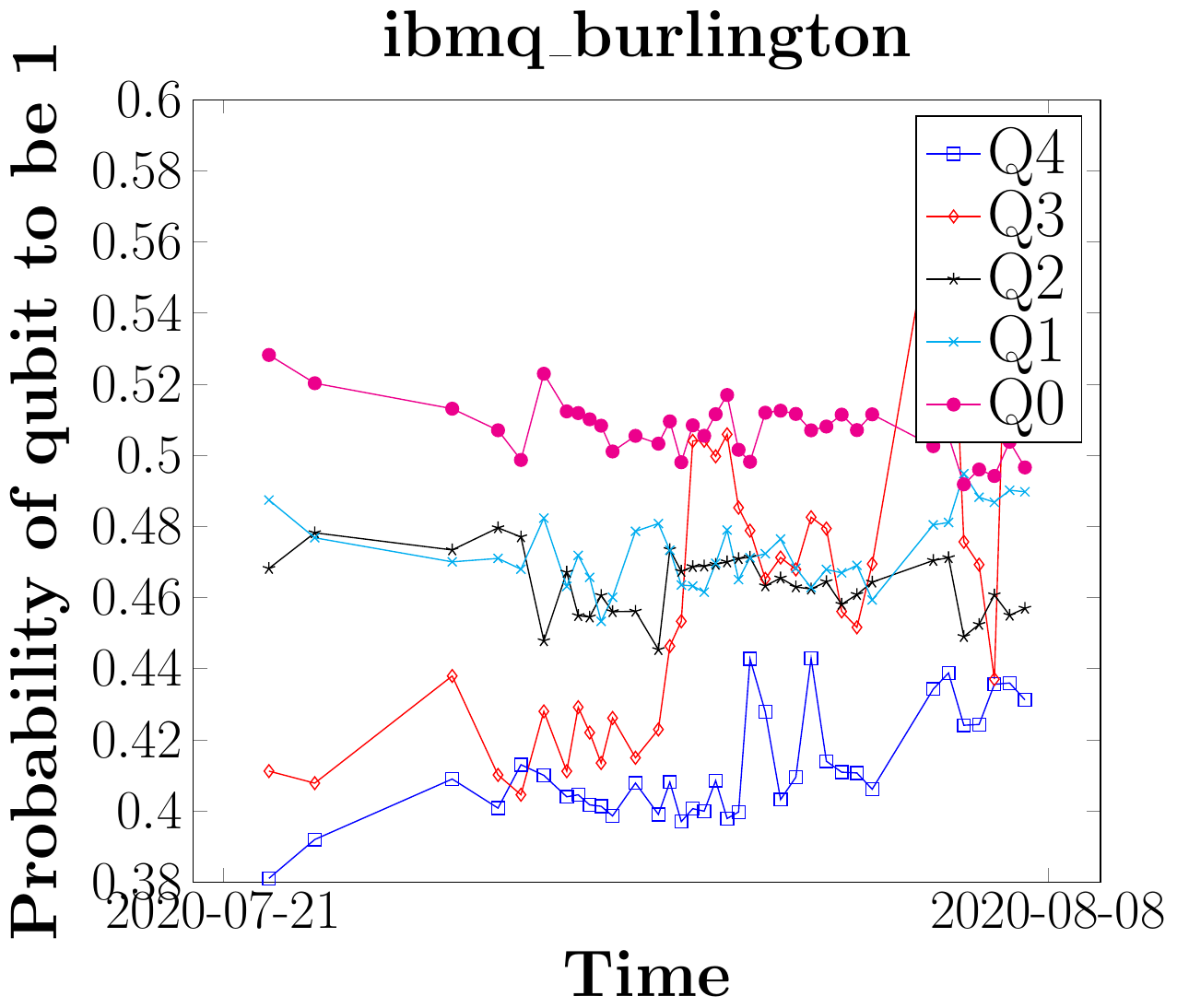}
        \caption{}
    \end{subfigure}
    
    \begin{subfigure}{0.31\linewidth}
        \includegraphics[width=\linewidth]{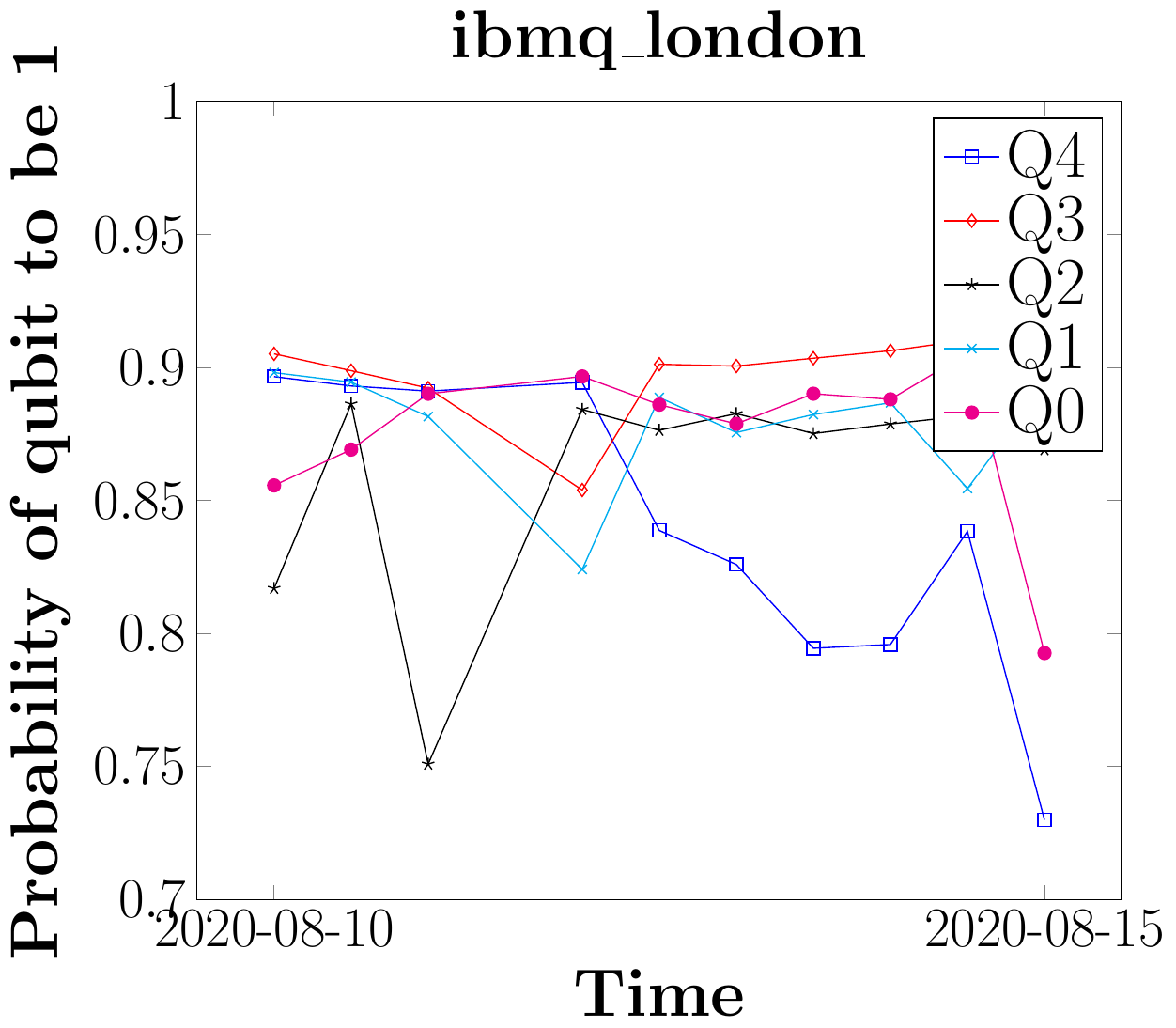}
        \caption{}
    \end{subfigure}
    \begin{subfigure}{0.31\linewidth}
        \includegraphics[width=\linewidth]{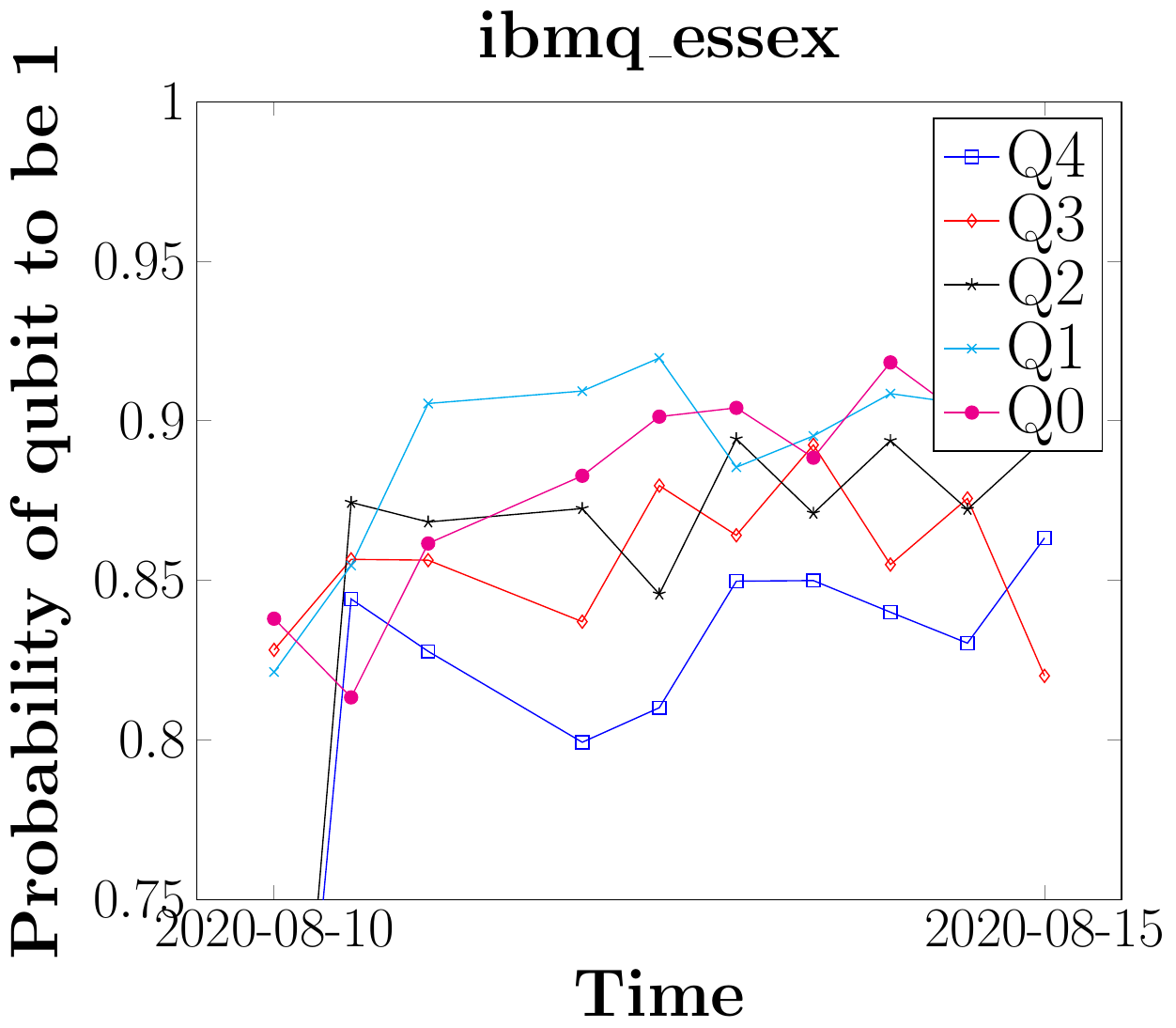}
        \caption{}
    \end{subfigure}
    \begin{subfigure}{0.31\linewidth}
        \includegraphics[width=\linewidth]{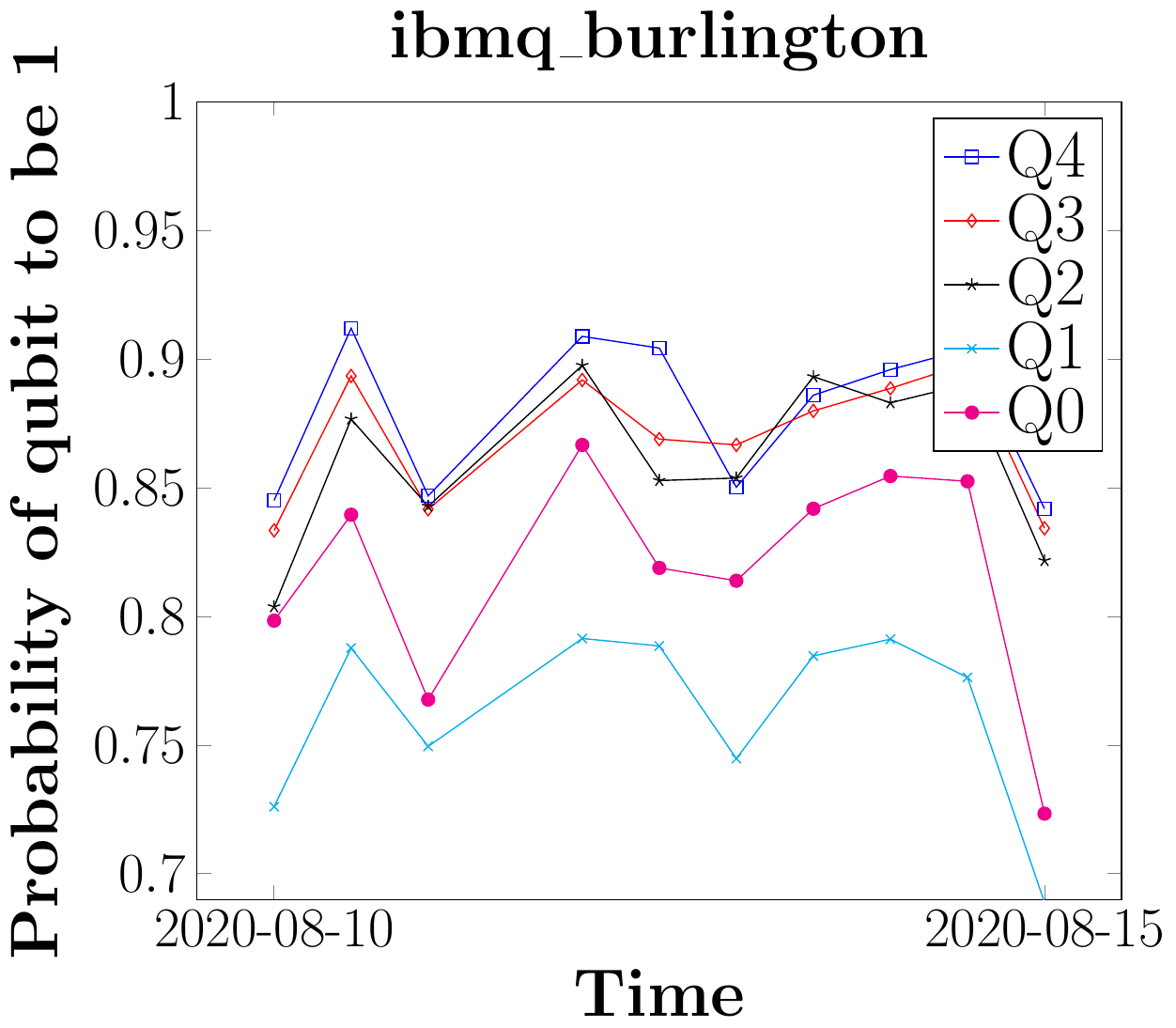}
        \caption{}
    \end{subfigure}
     \caption{Temporal variation in various quantum computers for the, (a)-(c) H-gate based; and, (d)-(f) decoherence-based QuPUF.\vspace{-0.1cm}}
   \label{fig:temporal_variation_plots}
\end{figure*}  

\textbf{Scenario-2) Low-fidelity allocation}: In this scenario, the number of qubits of both the target and the allocated quantum computers are the same (possibly, the structure as well). However, the qubit quality defined by error rates like CNOT error rate and single-qubit U2 error rate and decoherence/dephasing is worse for the allocated quantum computer. Such an allocation can reduce the program fidelity. 
If the user is running hybrid algorithms, the poor fidelity outcome can increase the convergence time (i.e., number of iterations). By freeing up the queue, the adversary will also be able to improve throughput.

\textbf{Scenario-3) Less-trusted quantum computers}: In future, less-trusted quantum computers could be available from 3rd parties that can allocate poor quality hardware and sabotage the output of the computing. Since the correct output of the optimization problem is not known, the user has to trust the sub-optimal result obtained from the quantum computer. In applications of national importance, this could have significant implication.

\textbf{Scenario-4) Rogue employee/malicious code in scheduler}: A rogue employee in trusted cloud vendor could try to sabotage the vendor's reputation by degrading the user compute fidelity just by tampering with the scheduling algorithm or rerouting the program to inferior hardware. Similar objectives can also be carried out if the scheduler is hacked by a malicious software. The rogue employee/malicious scheduler can also steal information by redirecting the programs to a 3rd party quantum hardware where they have full control. 



\subsection{Device Identification by QuPUF}
Fig. 3 shows the steps involved in QuPUF based device authentication. It itvolves following phases: 

\textbf{Registration:} 
First, a CRP database will be created similar to the conventional PUF (registration). For this step, the CRP of all qubits of each of the hardware will be collected and added to the CRP database. During validation the signature obtained from the hardware will be matched against the database for identification/validation.  

\textbf{Validation:} The QuPUF, which is a quantum circuit, will be sent as a workload to the quantum hardware (step-1). Through this workload, the aim is to obtain the measurement results to act as the device signature. The expectation is that each hardware will produce a unique device signature depending on the internal characteristics such as, qubit quality through error rates, number of qubits, coupling map of the qubits, etc. Once the signatures have been obtained for each device (step-2), the user will query the CRP database (step-3) which will provide the hardwware corresponding to the signature (step-4). The user can then validate if the hardware is same as expected/requested. 

\section{Quantum PUF and Experimental Results}
In this section, we explain the proposed QuPUFs and present experimental results. We also compute the QuPUF quality and
\begin{figure}[h!]
    \begin{subfigure}{0.5\linewidth}
        \includegraphics[width=\linewidth]{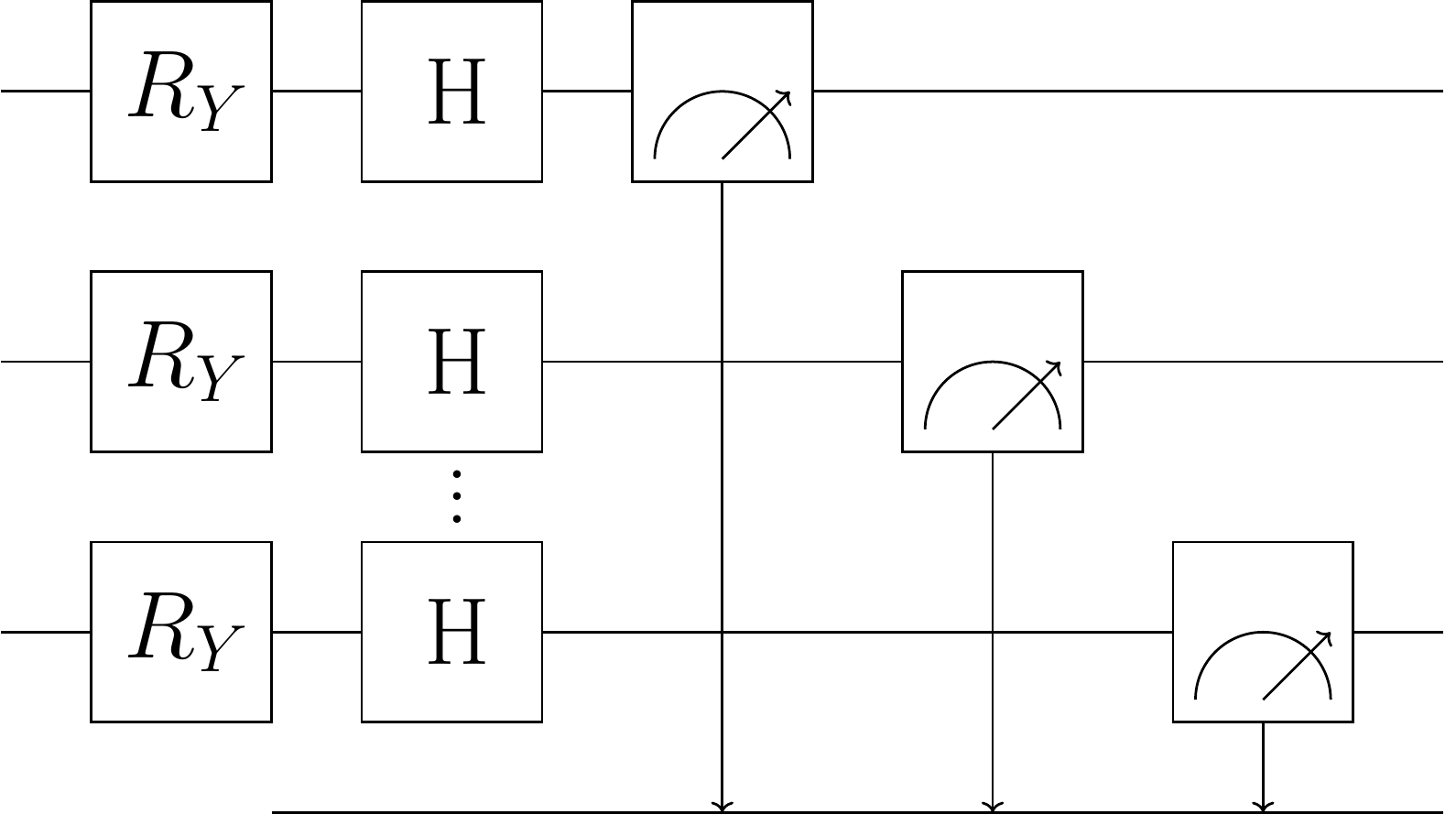}
        \caption{}
        \label{fig:f3a}
    \end{subfigure}
    \hspace{-0.3cm}
    \begin{subfigure}{0.5\linewidth}
        \includegraphics[width=\linewidth]{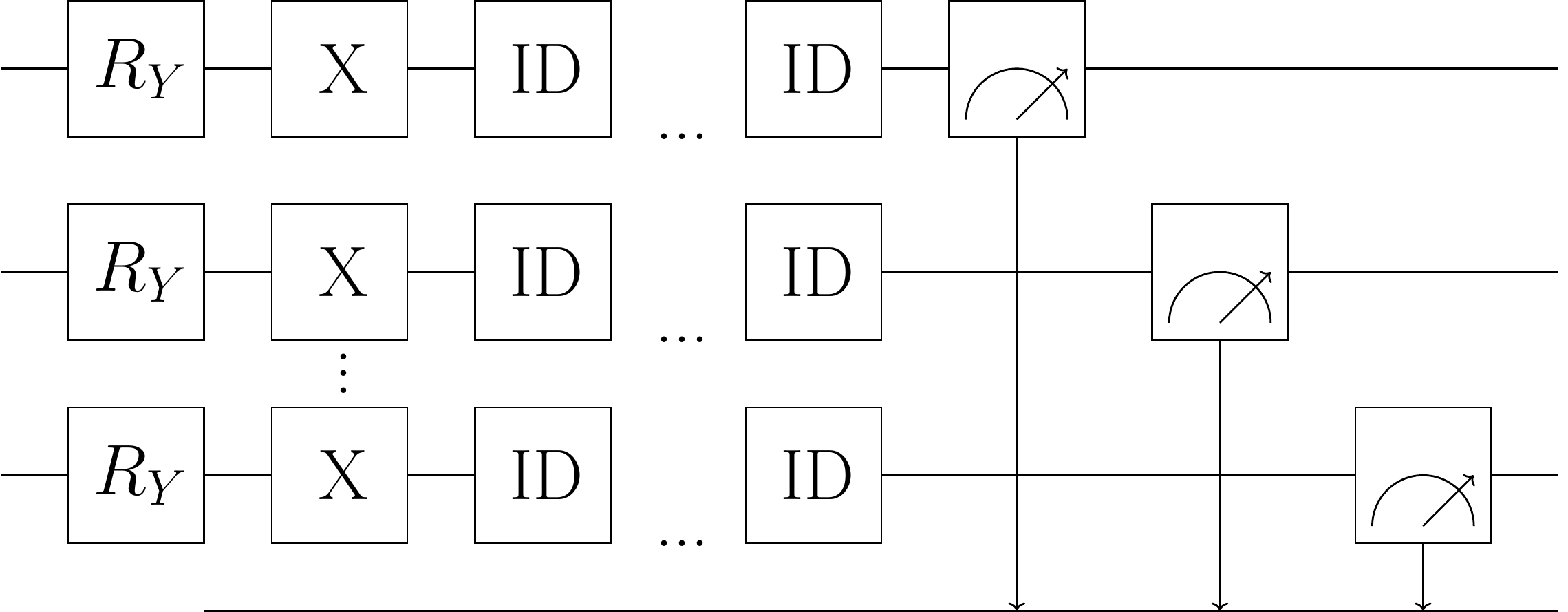}
        \caption{}
        \label{fig:f3b}
    \end{subfigure}
 \vspace{1em}
 \caption{Proposed QuPUFs: (a) Hadamard gate-based QuPUF; (b) decoherence-based QuPUF. The tunable rotation has been added for resilience.} \label{fig:f3}
 \vspace{-0.2em}
 \label{fig:qupufs}
\end{figure} \noindent improve them using parametric rotation.

\vspace{-1.5mm}
\subsection{Generic methodology and experimental setup}
Each qubit of the quantum hardware is distinct in terms of 1-qubit and 2-qubit gate error, readout error, decoherence/dephasing and crosstalk error rates. The general strategy of QuPUF design is to convert these error rates into a qubit signature which in turn, will form the hardware signature. A very naive QuPUF could just initialize the qubits to ground state and perform a readout. It will convert the readout error into a signature. This paper only exploits the 1-qubit gate error, readout error and decoherence error to generate the signature although other means of designing the QuPUF are also possible. The proposed basic QuPUFs have single challenge and response (i.e., weak PUF) whereas resilient QuPUF employ rotation as a challenge. It is possible to expand the CRP by adding more challenges. 
ibmq\_london, ibmq\_burlington and ibmq\_essex computers (Fig. \ref{fig:coupling_map} (a)) have been used for the basic QuPUFs. ibmq\_london has been used for resilient Hadamard gate-based QuPUF and ibmq\_vigo has been used for resilient decoherence-based QuPUF. For experiments with the QuPUFs, we have used real quantum hardware from IBM. For the basic QuPUFs, 75 experiments with 8192 shots per experiment were used, while for the resilient QuPUF, it was reduced to 20 experiments and 1024 shots per experiment due to long wait queue. The interval for measurements were also different for the basic and the resilient QuPUFs. 
\vspace{-1mm}
\begin{figure*} [!t] 
    \begin{subfigure}{0.24\linewidth}
        \includegraphics[width=\linewidth]{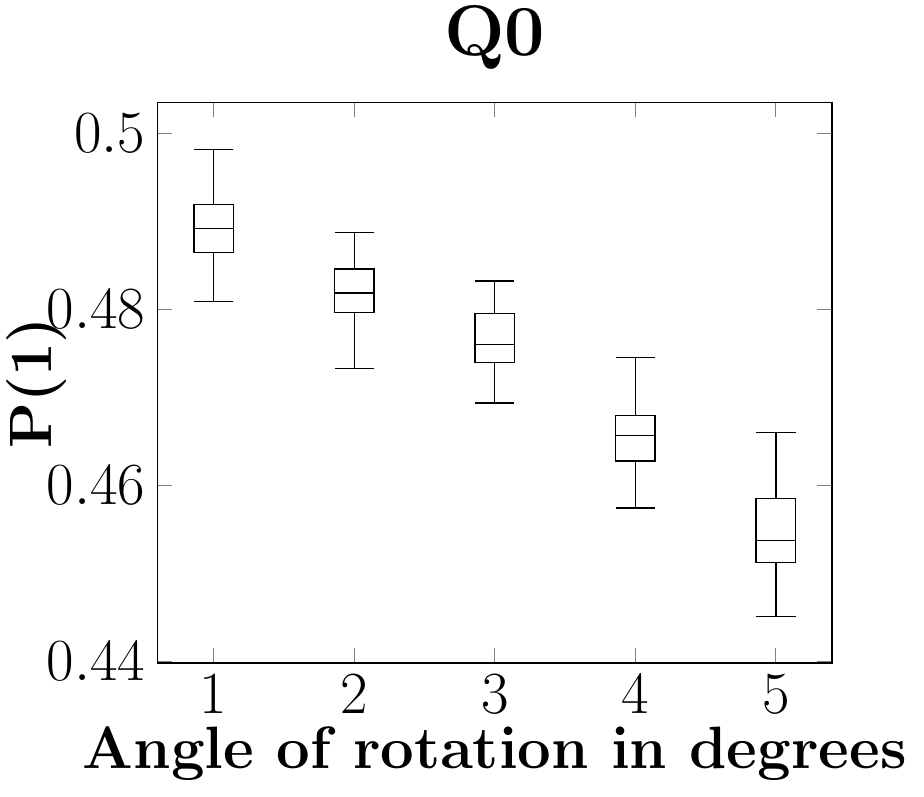}
        \caption{}
    \end{subfigure}
    \begin{subfigure}{0.24\linewidth}
        \includegraphics[width=\linewidth]{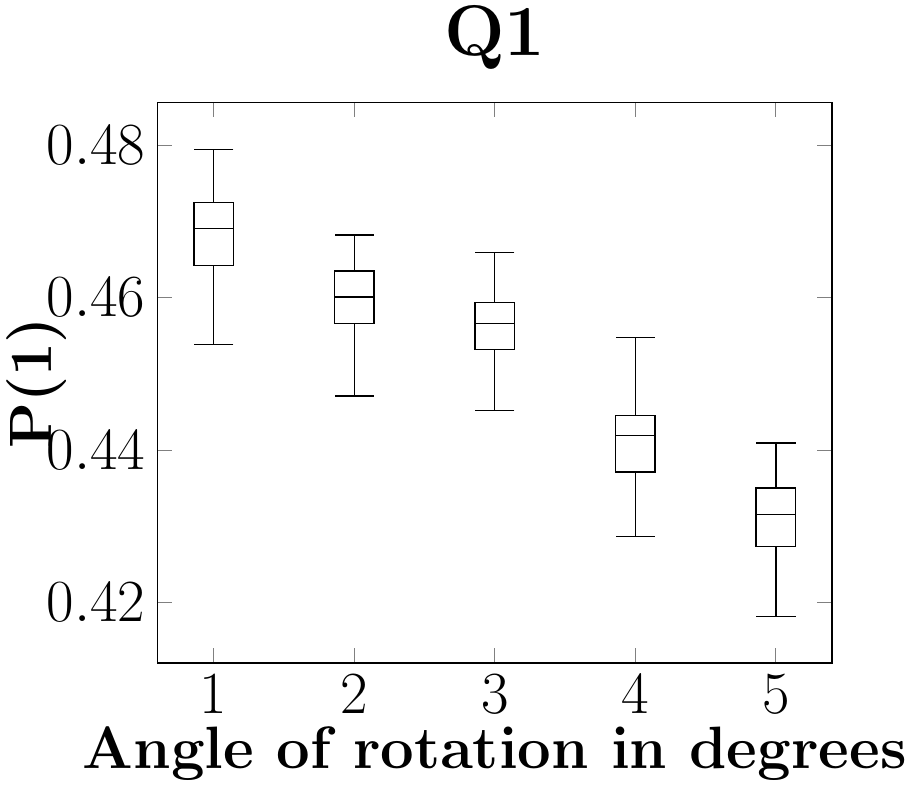}
        \caption{}
    \end{subfigure}
    \begin{subfigure}{0.24\linewidth}
        \includegraphics[width=\linewidth]{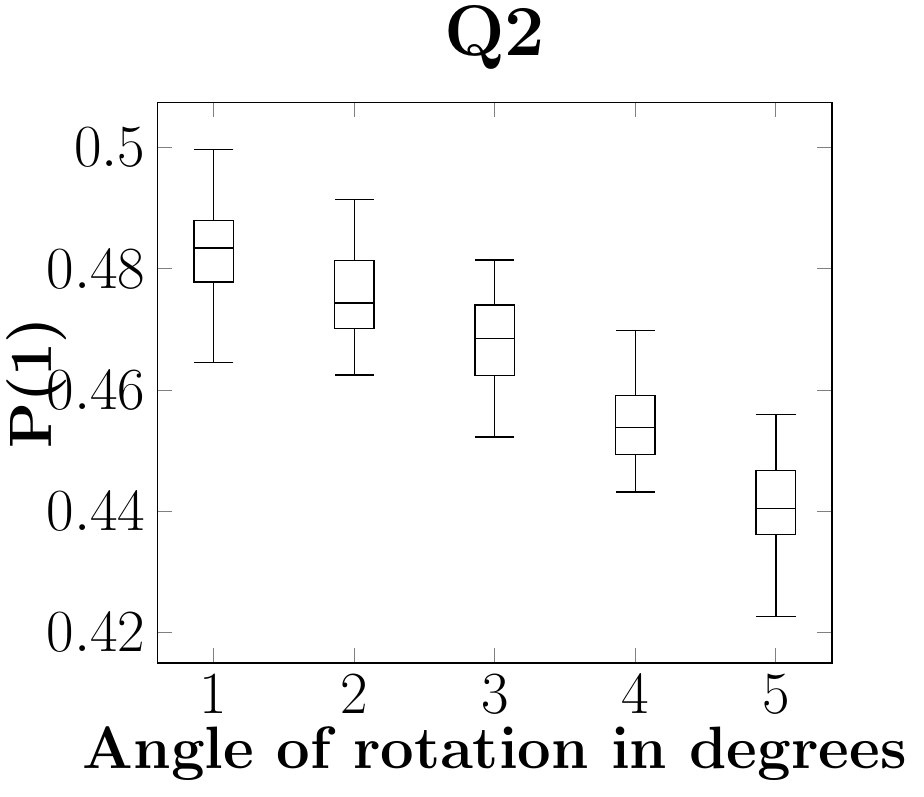}
        \caption{}
    \end{subfigure}
    \begin{subfigure}{0.24\linewidth}
        \includegraphics[width=\linewidth]{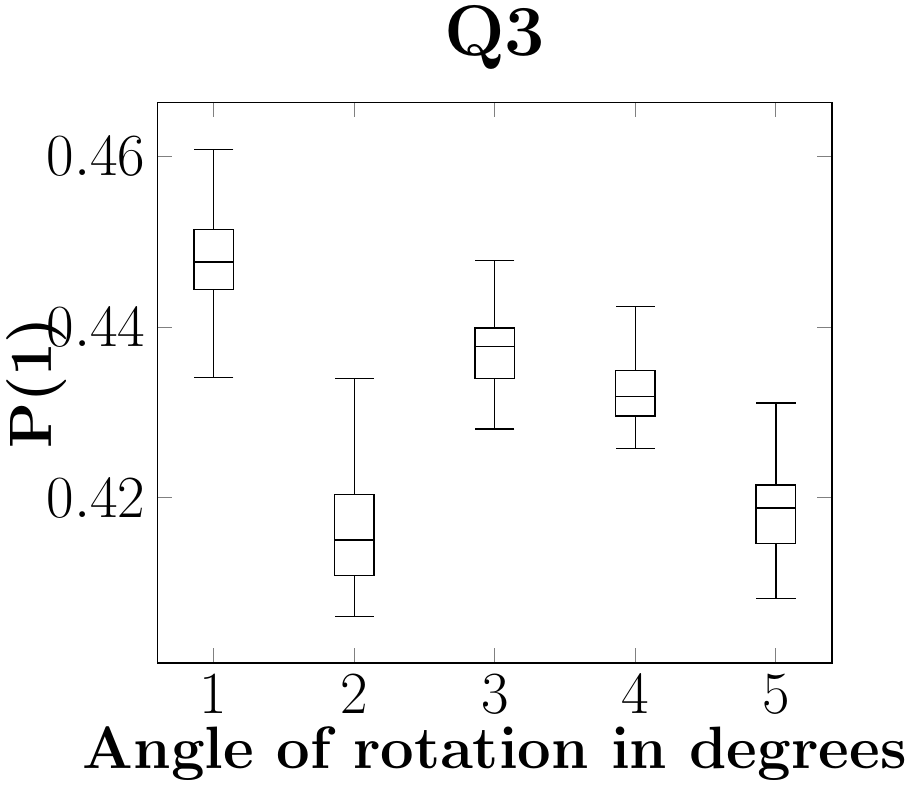}
        \caption{}
    \end{subfigure}
    \begin{subfigure}{0.24\linewidth}
        \includegraphics[width=\linewidth]{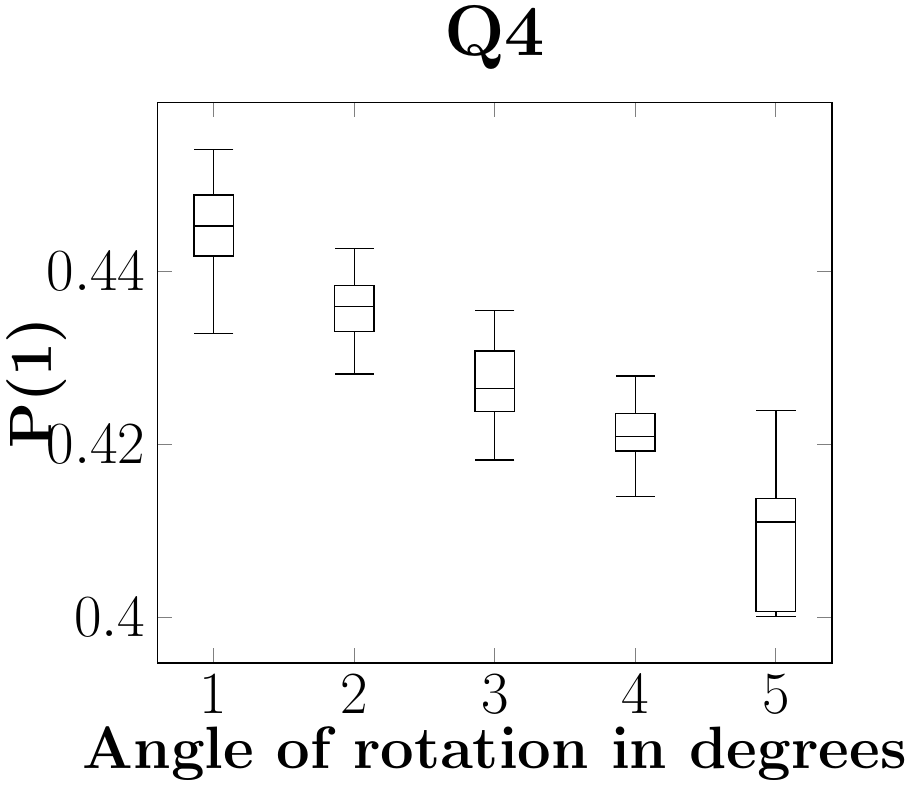}
        \caption{}
    \end{subfigure}
    \begin{subfigure}{0.24\linewidth}
        \includegraphics[width=\linewidth]{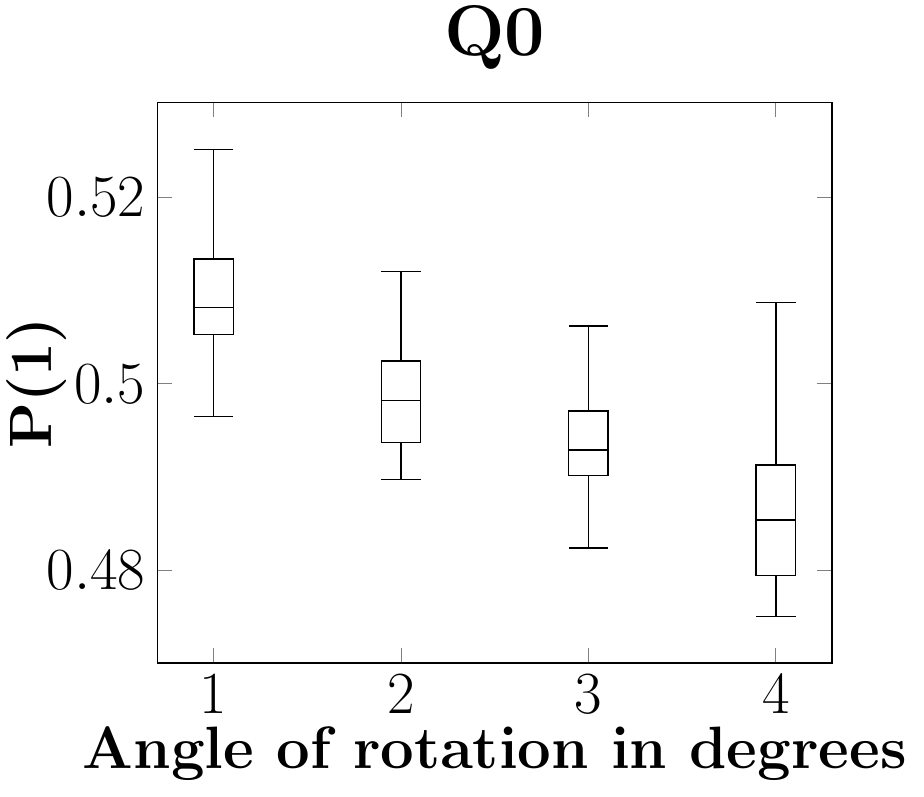}
        \caption{}
    \end{subfigure}
    \begin{subfigure}{0.24\linewidth}
        \includegraphics[width=\linewidth]{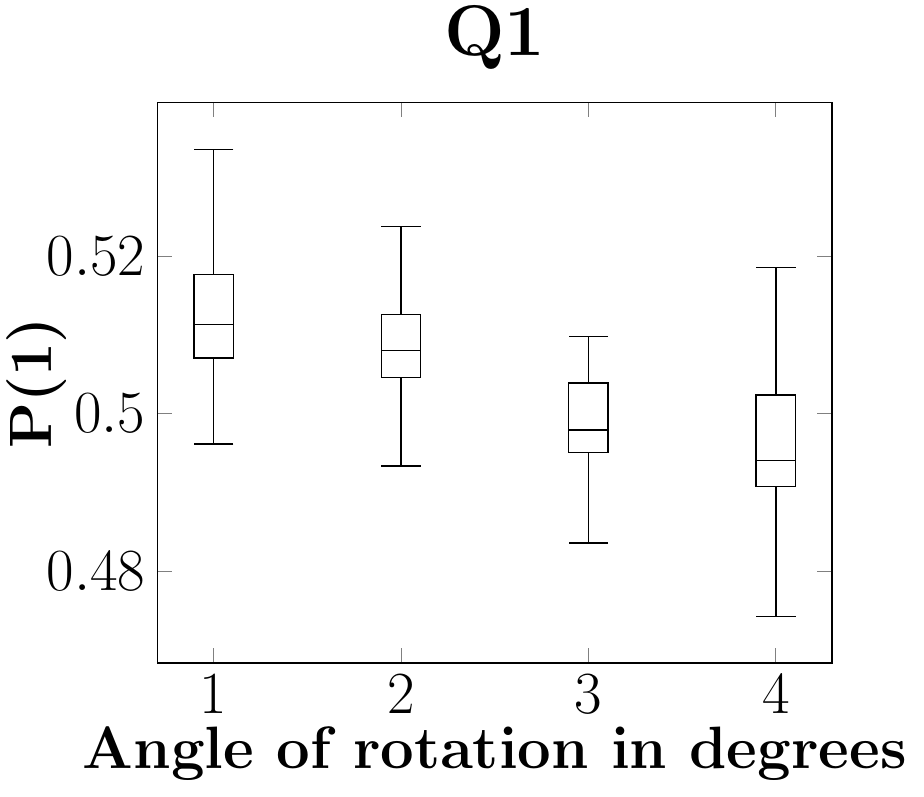}
        \caption{}
    \end{subfigure}
    \begin{subfigure}{0.24\linewidth}
        \includegraphics[width=\linewidth]{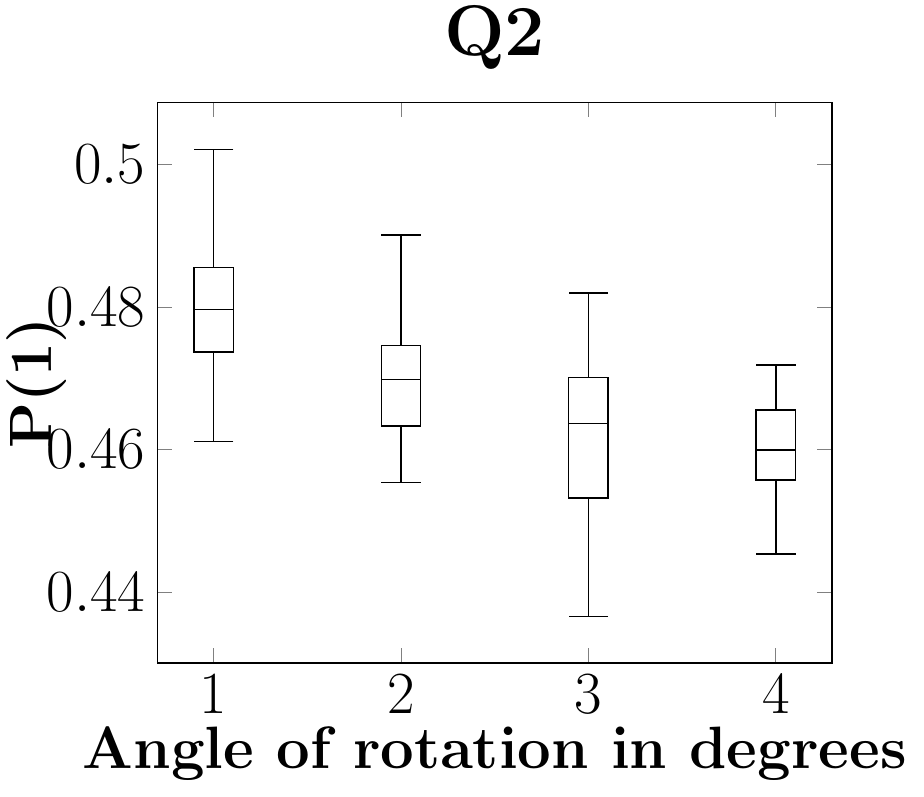}
        \caption{}
    \end{subfigure}
    \begin{subfigure}{0.24\linewidth}
        \includegraphics[width=\linewidth]{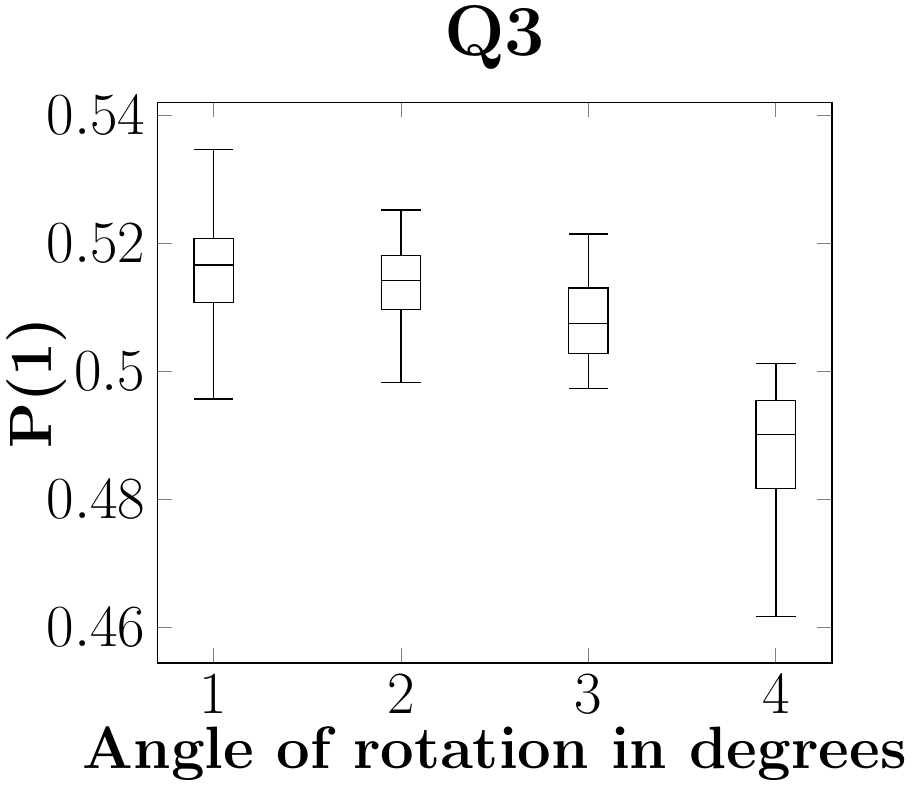}
        \caption{}
    \end{subfigure}
    \begin{subfigure}{0.24\linewidth}
        \includegraphics[width=\linewidth]{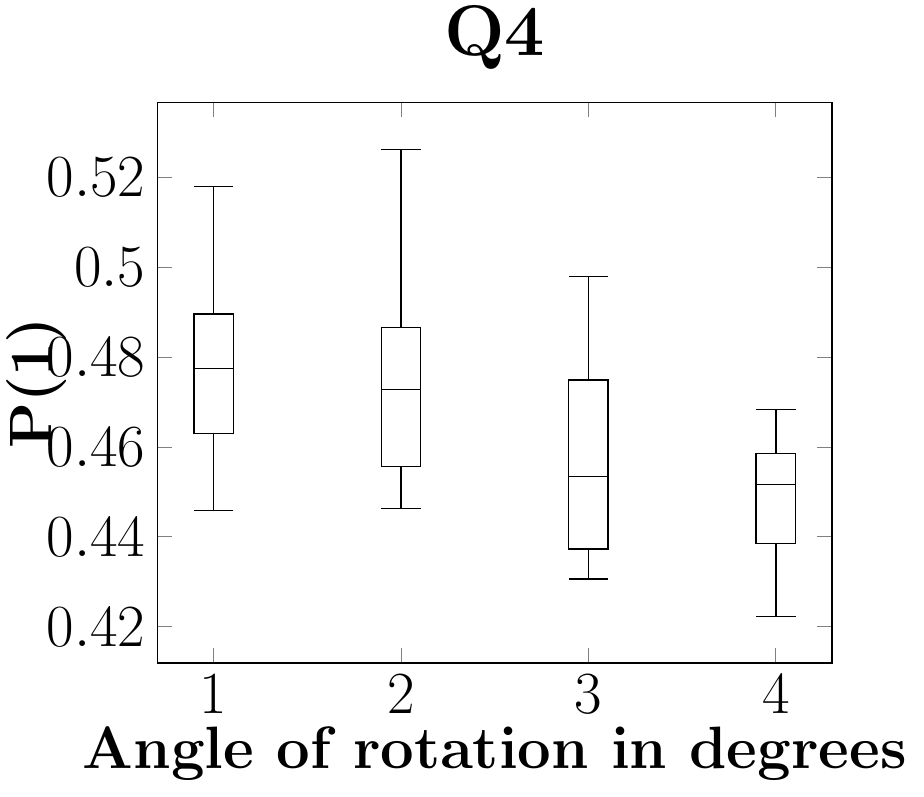}
        \caption{}
    \end{subfigure}
    \begin{subfigure}{0.24\linewidth}
        \includegraphics[width=\linewidth]{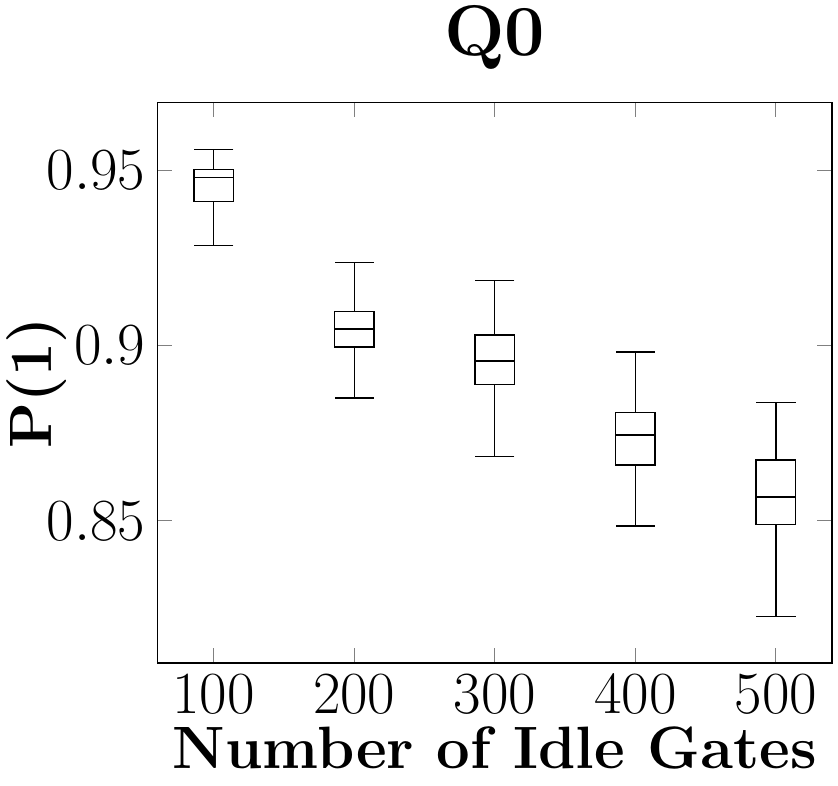}
        \caption{}
    \end{subfigure}  
    \begin{subfigure}{0.24\linewidth}
        \includegraphics[width=\linewidth]{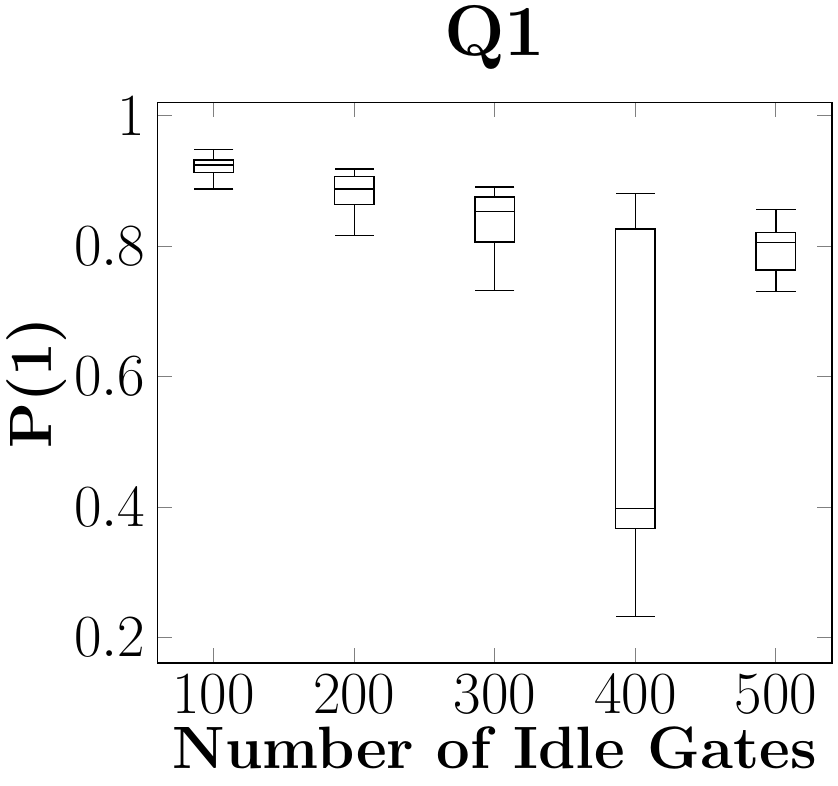}
        \caption{}
    \end{subfigure}
    
    \begin{center}
      \begin{subfigure}{0.24\linewidth}
        \includegraphics[width=\linewidth]{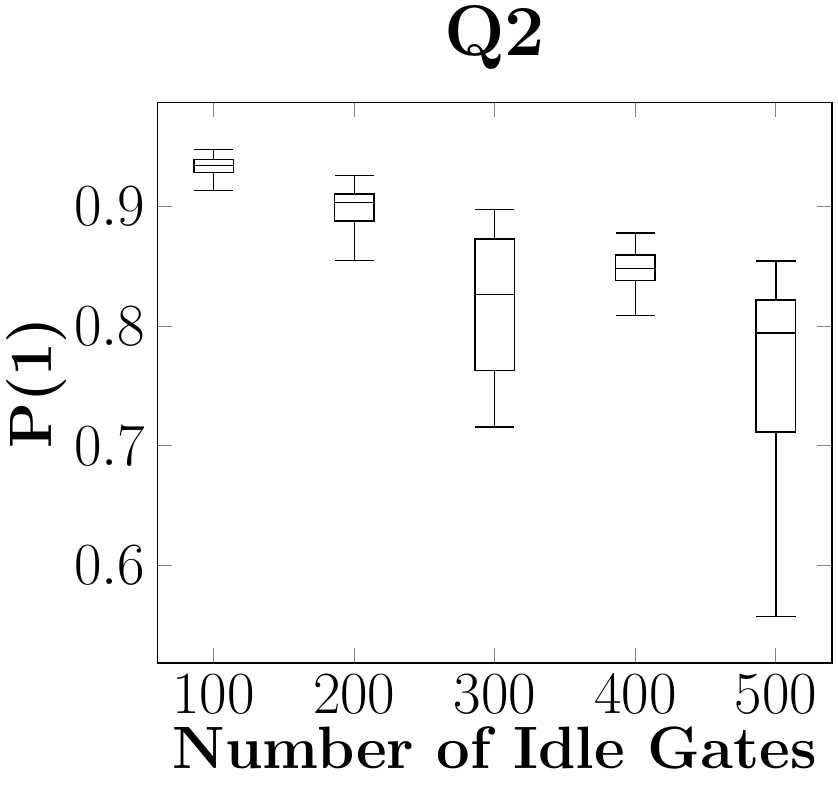}
        \caption{}
    \end{subfigure}
    \begin{subfigure}{0.24\linewidth}
        \includegraphics[width=\linewidth]{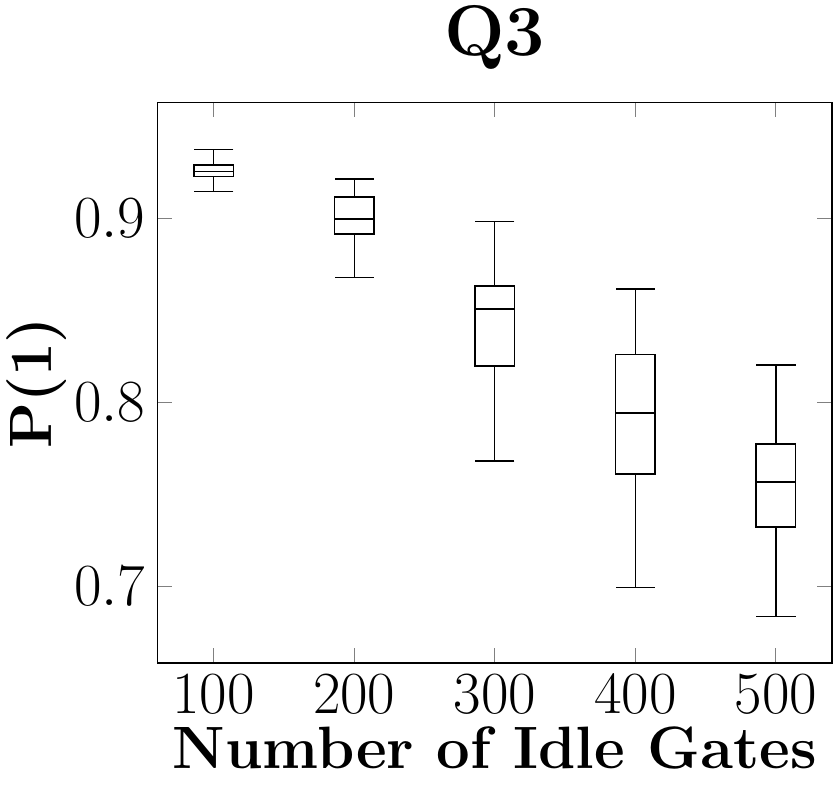}
        \caption{}
    \end{subfigure}
    \begin{subfigure}{0.24\linewidth}
        \includegraphics[width=\linewidth]{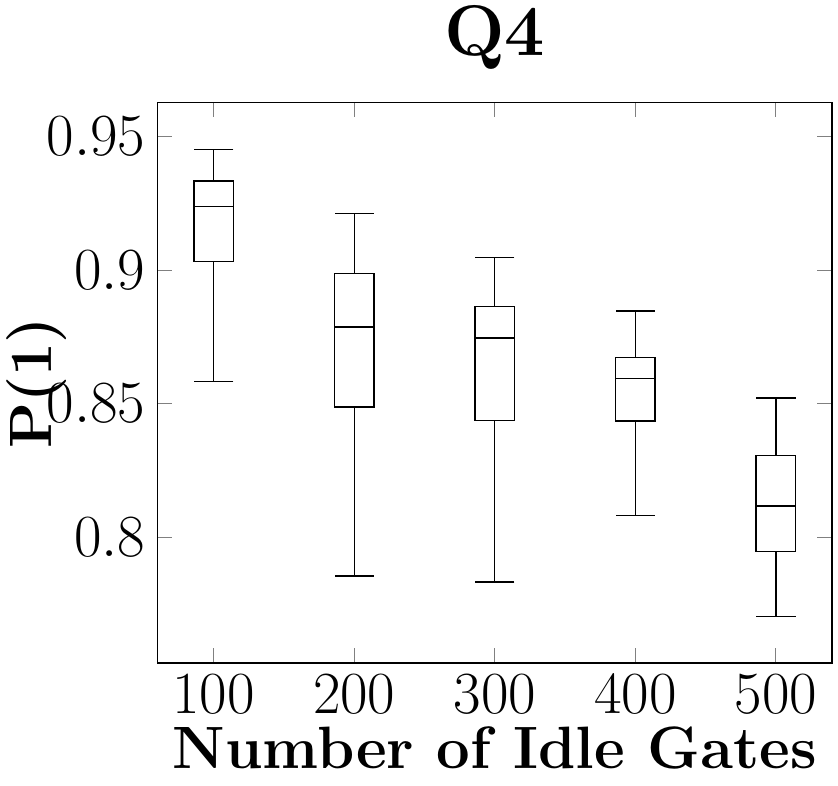}
        \caption{}
    \end{subfigure}  
    \end{center}

 \caption{Box plot of various qubits with different number of idle gates for \textbf{ibmq\_london ((a)-(e)), ibmq\_essex ((f)-(j)) and ibmq\_vigo ((k)-(o))}. The sigma of the distribution minimizes with a specific number of idle gates.} \label{fig:boxplots}
\end{figure*}
\subsection{Basic QuPUFs}
\textbf{Hadamard gate-based QuPUF:}
This QuPUF exploits the biasing of the qubits towards 1 or 0 state to generate the response. The biasing could be a result of readout error (typically large) or the gate error (small for single qubit gates). Each qubit is initialized to zero state at the beginning. Next, the qubits are placed in a superposition state (using a Hadamard gate) followed by the measurement (Fig. \ref{fig:qupufs}(a)). Ideally, the qubits should produce equal probability of both 0 and 1 states. However, the probability is expected to be biased towards either zero or one, depending on the errors that will act as unique device signature. 

\textbf{Decoherence-based QuPUF:} 
This QuPUF exploits the differences in the decoherence times of the qubits to generate the response. The qubits are placed in an excited state and allowed to decohere for a fixed amount of time followed by the measurement operation. Some qubits decohere more and exhibit more 0 than 1 and vice versa is true for qubits that decohere less. The probability of 1 state acts as the unique response. We first initialize the qubits to ground state. Next, we flip the state of the qubit using the X-Gate from 0 to 1 state. Finally, we allow it to decohere back to zero state by keeping the circuit idle i.e., by using idle gates followed by measurement (Fig. \ref{fig:qupufs}(b)). 

\textbf{Experimental results:}
Fig \ref{fig:temporal_variation_plots} (a)-(c) shows the probability of 1 of the H-gate based QuPUF for the three hardware artifacts collected over few days. It can be observed that the probabilities of each qubit for all the quantum computers fluctuate significantly i.e., the device signature is sensitive to temporal variations. However, the signature also exhibits spatial variation which enables us to distinguish each qubit via the probability of 1. Fig. \ref{fig:temporal_variation_plots} (d)-(f) depicts the results of the three quantum computers for the decoherence-based QuPUF. Here, the spatial variation is more distinguished in terms of the mean separation, and the temporal variation is also relatively lower compared to the H-gate QuPUF. However, none of the qubits decohered to ground state implying that full decoherence did not occur due to less number of idle gates.

\textbf{Interpretation of the results:}
Fig \ref{fig:temporal_variation_plots} shows temporal variation of probability of 1 for various qubits in each of the three hardware for Hadamard gate-based and decoherence based PUFs. It clearly shows that absolute value of probability of 1 may not provide clear PUF signature. However, analytical techniques such as, mean and standard deviation of the PUF response can be used reliably. Fig \ref{fig:boxplots} shows the boxplots for the resilient QuPUFs with varying rotation angle and varying number of idle gates. Here, the criteria for selection is to choose angle/ number of idle gates for every qubit which provide as much inter-qubit mean separation as possible, and also the least intra-qubit standard deviation.

For example, for Q0, we can select 4° and for Q1, we can select 5°. By selecting these values, we ensure that there is a mean separation (~0.47 for Q0 and ~0.43 for Q1) and the deviation is relatively less compared to other angles, implying that the variation of the selected angles is less.

\vspace{-1mm}
\subsection{Resilient QuPUFs}

\textbf{QuPUF with tunable rotation:}
The temporal variation of the quantum circuit is a function of the quantum state. It has been noted that adding parametric gate to the quantum circuit and tuning could optimize the resilience to dynamic variation \cite{alam2019addressing}. Following similar line of thought, we added a tunable rotation gate (e.g., $R_Y$ gate) to the H-gate and the X-gate in each qubit for resilience of the proposed QuPUFs to temporal variation. The rotation angle could be varied slightly e.g., from $1^{\circ}$ to $5^{\circ}$ (Fig. \ref{fig:qupufs}).  
The tunable rotation can also act as a challenge to increase the challenge-response pair (CRP). A rotation towards 0 is expected to shift the probability of 1 towards the ground state. However, the inherent bias remains present for each quantum computer serving as a unique signature. For the decoherence based QuPUF, the number of idle gates has been varied to study the impact on stability for a fixed rotation angle. The rotation angle and the number of idle gates providing the optimal values of inter- and intra-HD are selected for the QuPUF (experimental HD results are described in Section~\ref{subsec:hd}).

\textbf{Experimental results:}
We sweep the rotation angle of H-gate based QuPUF and plot the mean and sigma of the probability of 1 for each qubit in Fig. \ref{fig:boxplots}. 
For the decoherence QuPUF, we sweep the number of idle gates from 100-500. 
Fig.~\ref{fig:boxplots} (k) - (o) shows the box-plots for each qubit with variable number of idle gates.
\vspace{-1mm}
\subsection{Quality evaluation of the QuPUFs}\label{subsec:hd}
\begin{figure}
    \centering
    \includegraphics[width=0.95\columnwidth]{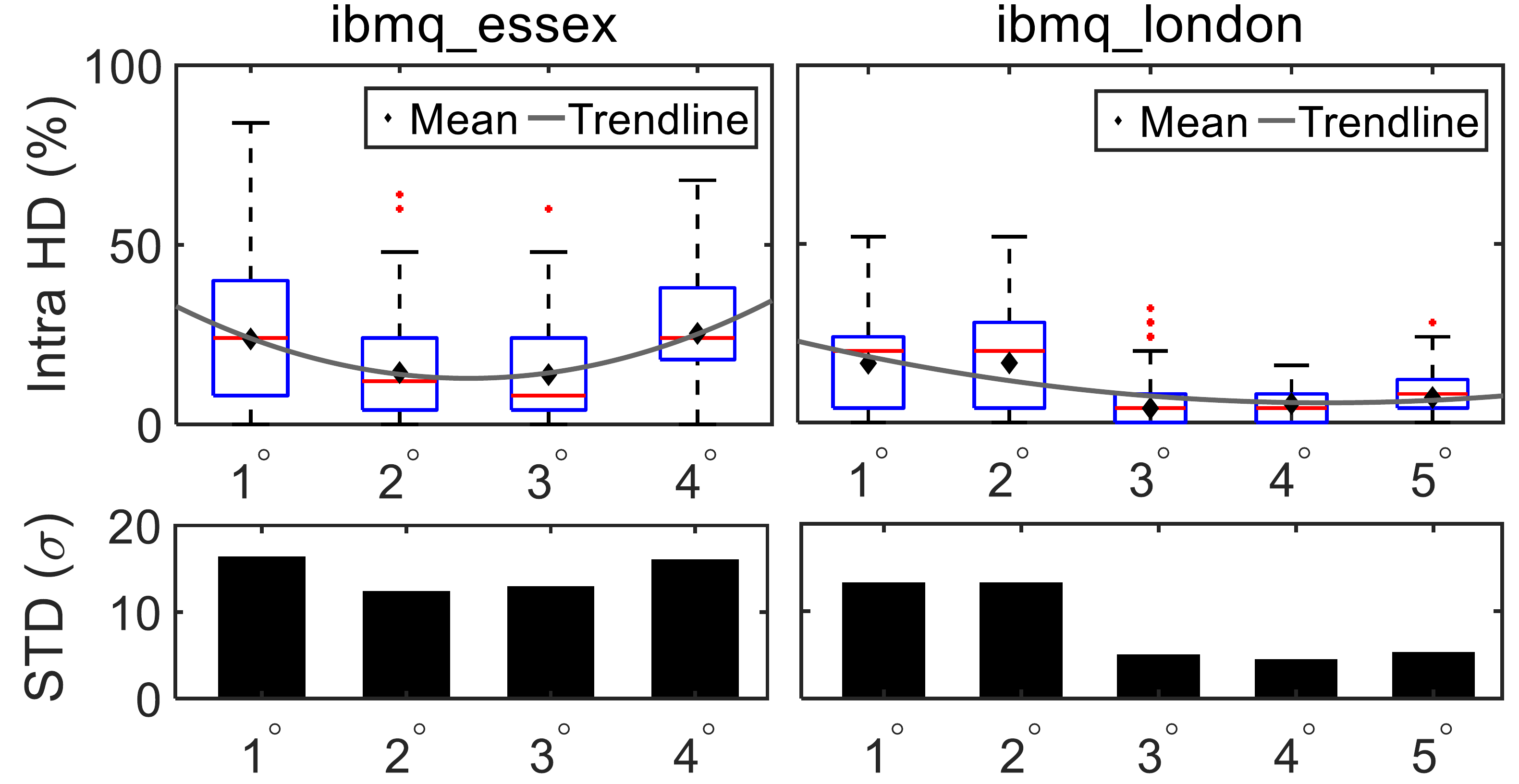}
    \caption{Intra HD from two backends ibmq\_essex and ibmq\_london. X-axis shows the rotation angle $\theta$ in $R_Y(\theta)$ for H-gate based PUF. The trend line shows as we vary the angle, intra-HD also varies for an optimal angle it is lowest. For both ibmq\_essex and ibmq\_london, $3^\circ$ shows the minimum intra-HD.\vspace{-0.1cm}}
    \label{fig:intra-hd}
\end{figure}
\textbf{HD calculation:} For calculating the intra- and inter-HD, we need to convert the analog value of hardware signature i.e., probability of `1' into a digital form. We split the range of probabilities in 32 steps for a 5-bit signature per qubit. Thus, we get a 25 bit (5 qubit $\times$ 5 bits/qubit) signature for each data point for each of the 5 qubit hardware. 

To compute intra-HD, we calculate the HD between all pairs of data points for the same quantum computer obtained over time and take the mean. To estimate the inter-HD, we compute the HD between all pairs of data-points for two different quantum computers and take the mean. The absolute value of the inter- and intra-HD will depend on the precision of the signature.
We sweep the signature precision from 4-bit to 9-bit to choose the optimal value.

\textbf{Experimental results:} Fig.~\ref{fig:intra-hd} shows the intra-HD distributions from ibmq\_essex and ibmq\_london for 5-bits precision (H-gate based PUF with rotation). The top traces are the box-plots of distribution with explicitly plotted trend-lines of the means. The bottom traces show the standard deviations $(\sigma)$. The plots depict that the intra-HDs vary with rotation angle $(RY(\theta))$ and exhibits an optima. For both ibmq\_essex and ibmq\_london, $3^\circ$ is the optimal angle with lowest intra-HD (ibmq\_essex $13.82\%$ and ibmq\_london $3.94\%$). 

\begin{figure}
    \centering
    \includegraphics[width=1.0\columnwidth]{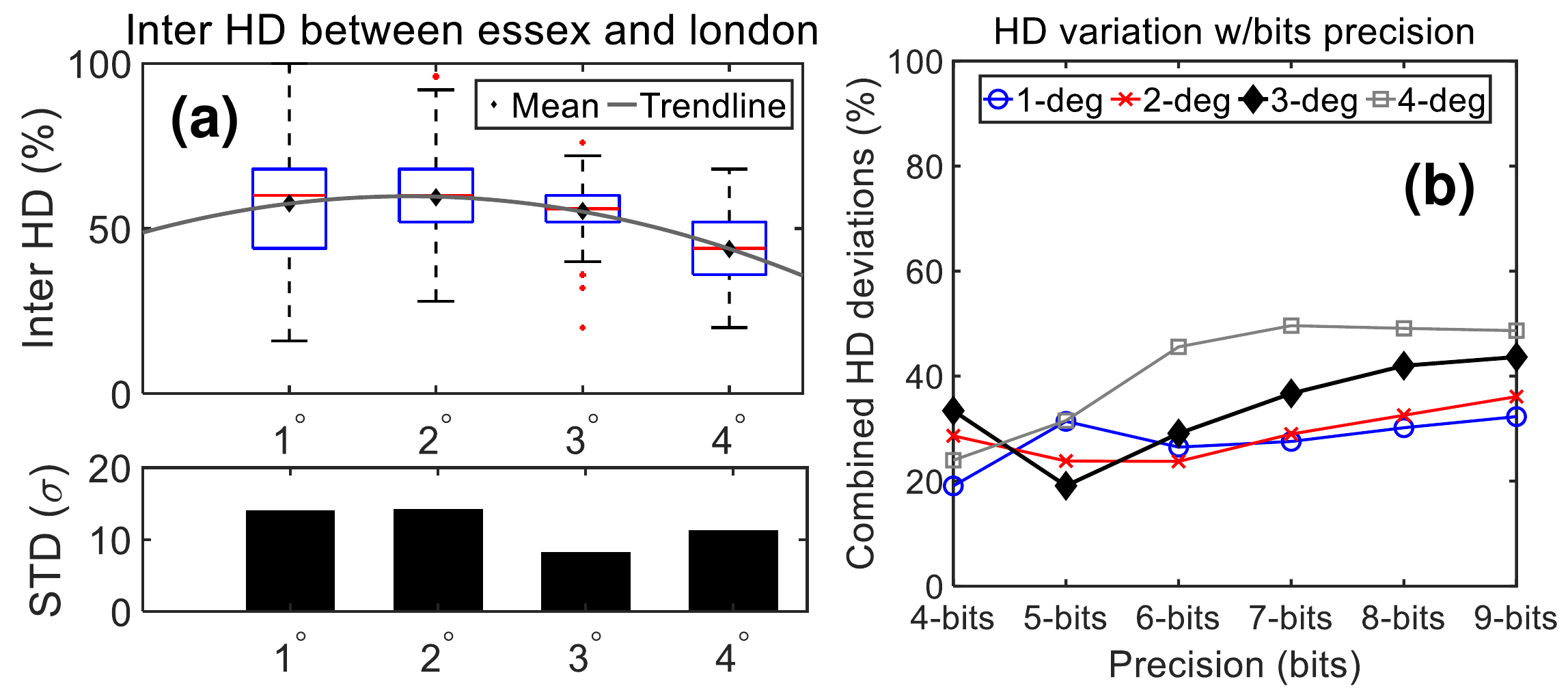}
    \caption{(a) Inter-HD between ibmq\_essex and ibmq\_london; (b) combined inter- and intra-HD with varying bit precisions.\vspace{-0.1cm}}
    \label{fig:combined-hd}
\end{figure}

Fig.~\ref{fig:combined-hd}(a) shows the inter-HD between ibmq\_essex and ibmq\_london for 5 bits precision. This plot also shows inter-HD varies with rotation angle, and it is optimal for $3^\circ$ ($55.3\%$). 

Finally, we plot both inter- and intra-HD with bit precisions and rotation angles (Fig.~\ref{fig:combined-hd}(b)). As we want inter-HD to be close to $50\%$  and intra-HD close to $0\%$, a \emph{combined HD deviation} metric is defined as $|inter\_HD - 50| + |intra\_HD - 0|$ which captures the deviations in both inter- and intra-HD. The lower combined value is desirable. We plot the results for inter-HD for ibmq\_essex and ibmq\_london and intra-HD from ibmq\_essex. As we vary bit precision, the combined metric gives the optimal value for $3^\circ$ rotation and $5-$bits precision (optimal combined value $19.12\%$). Thus, setting angle and precision to $3^\circ$ and $5$-bits are best for QuPUF. Results with intra-HD from ibmq\_london also shows similar behavior, and therefore, omitted for brevity.

\textbf{Comparison between QuPUFs:} We also compute intra-HD for decoherence based PUF with data collected from ibmq\_vigo with variable number of idle gates (100 -- 400 idles). The intra-HD varies from $13\%$ (100 idles) -- $27\%$ (400 idles) with a 5-bit precision. Therefore, a lower number of idle gate is better in decoherence-based PUFs in terms of intra-HD. Overall, H-gate based QuPUF performs better in terms of stability. 

\section{Discussions and Limitations}
This section describes various aspects of the proposed QuPUF and potential limitations.


\textbf{Ensuring trust at hardware level:}
Validity and quality of cloud-based hardware providers is an important factor while establishing trust between the user and the hardware. However, it is also necessary to establish trust on hardware level itself. This is primarily because, (i)  quantum hardware can provide high-quality signature like conventional CMOS PUFs and, (ii) it is ultimately the quality of the hardware that determines the accuracy of the result. An inferior hardware allocation can lead to higher costs and poor-quality solution which is undesirable. Hence, there is need of an assurance that the desired hardware is allocated to the user.

\textbf{Bypassing QuPUF-based validation:} It is possible that the QuPUF circuit is identified by the scheduler and routed to the correct hardware however, the actual user workload is rerouted to the incorrect hardware. This is possible if the vendor side scheduler is aware of the existence of QuPUF and employs a detection routine. This scenario can be addressed by embedding the QuPUF within the user workload. For example, user can validate the identity of few qubits while the other qubits are used for computation. This is possible for large workload running on large quantum hardware. For example, 3-4 qubits can be used to validate the identity of a 23-qubit hardware while the remaining qubits are used for computation. This approach will reduce the number of compute qubits. One can also use uncomputation to free up few qubits that have completed computation early and use them to run QuPUF circuit at no added overhead ~\cite{ding2020square}.   

\textbf{Other QuPUF designs: }
Specific QuPUFs to exploit to readout error, 2-qubit gate errors and crosstalk can also be designed and evaluated for stability and uniqueness. It is also possible combine the responses of various QuPUFs to enhance the quality. For example, the response of H-gate and decoherence-based QuPUFs can be combined to identify the hardware more accurately than using them in isolation. \vspace{-0.1mm}

\textbf{Challenge-Response Pairs (CRP): }
The present implementation of the QuPUF is a weak PUF with one rotation gate and one rotation angle for each qubit. This can be expanded further by adding additional rotational gates to the already present $R_Y$ gate, so that the challenge depends on more than one rotation angle. In such a scenario, the angle of rotation for each rotation gate can be considered a challenge. This approach will provide exponential CRP, with linear number of rotational gates (with rotations that provide stable HD).

\textbf{Decohorence vs \#Idle Gates: }
The intended effect of decoherence was not observed because the idle time produced by the idle gates was less than the coherence time (relaxation) of the hardware. The idle gates' purpose is to pass time in order to allow the qubits to relax to ground state. Since lesser number of idle gates were present in the circuit, the qubits did not get enough relaxation time, and as a result, did not decohere. This can be resolved by adding a greater number of idle gates. This requirement for higher number of idle gates was not supported until very recently by IBM systems.

\textbf{Vulnerability to temporal variation: }
As seen from Fig. \ref{fig:temporal_variation_plots}, the QuPUFs are sensitive to temporal variation. However, the intra-HD and inter-HD values obtained are satisfactory. This implies that even if one obtains dynamic hardware signatures, they can be identified since they are spaced out with respect to intra-HD and inter-HD. Also, the current trend shows that quantum industry is able to reduce the noise levels and increase the decoherence time aggressively. Therefore, the effectiveness of the proposed PUFs is expected to improve in future.

\textbf{Unstable decoherence rates:}
The decoherence rates of modern NISQ computers are unstable, which poses a challenge for decoherence-based QuPUF. Varying decoherence rates will give varying amount of decoherence, and this will be reflected in the output. This might also cause increased readout error as measuring a qubit takes significantly longer than unitary operations on qubits, and during measurement, the qubits being measured may change their states due to decoherence \cite{leymann2020bitter}. Nevertheless, decoherence-based PUF is a potential direction to identify a quantum hardware once the variations are controlled at the hardware level.

\textbf{Other applications of the QuPUFs:} The proposed QuPUFs can also be used to address other security challenges such as, Man-In-The-Middle (MITM) Attack. If the attacker tampers with the device signature it will be detected during the signature verification stage. The QuPUF signature can also be used for non-repudiation of data. For this application, the QuPUF signature will be appended with the results of a computation from a quantum computer to authenticate the computation outcome. 

\textbf{Comparison with existing remote attestation protocols:}
Note that the proposed QuPUF is a quantum-hardware security primitive that can be used as a building block of a security protocol (e.g., Intel SGX/TPM) to establish trust between - the user and the service provider - in a quantum-cloud computing environment. For instance, Intel SGX performs remote device attestation using Enhanced Privacy ID (EPID). EPID consists of the following four elements: member private key, group public key, message to be signed, and signature revocation proof list. In an SGX-like security platform for quantum-cloud, QuPUFs can be used to generate unique private keys for the quantum hardware (members). Although to the best of our knowledge, such security protocols are not in use today in the quantum-cloud computing platforms, we expect to see developments in this domain soon.


 \vspace{-0.2cm}
\section{Conclusion}
We proposed two flavors of QuPUFs to establish trust in the public cloud-based quantum hardware. The proposed QuPUFs are thoroughly analyzed for uniqueness and stability on real quantum hardware. Our study indicated that minor tuning of parametric rotation of the QuPUF and choice of bit precision of the signature can optimize the response in presence of temporal and spatial variation in qubit quality. Experiments on real IBM quantum hardware show that the proposed QuPUF can achieve inter-die HD of 55\% and intra-HD as low as 4\%. The proposed QuPUFs can address wide range of security and trust issues associated with quantum computing. 

As time progresses, the effectiveness of the proposed PUFs will improve due to sophisticated quantum control and temporal error mitigation efforts employed by quantum computing industry to improve the quality of the hardware.

\textbf{Acknowledgements:} The work is supported in parts by National Science Foundation (NSF) (CNS-1722557, CCF-1718474, OIA-2040667, DGE-1723687 and DGE-1821766) and seed grants from Penn State Institute for Computational and Data Sciences and Penn State Huck Institute of the Life Sciences. We acknowledge the use of IBM Quantum Services for this work. The views expressed are those of the authors, and do not reflect the official policy or position of IBM or the IBM Quantum team. \cite{IBMQuantum}


\vspace{-4cm}

\begin{IEEEbiography}[{\includegraphics[width=1in,height=1.25in,clip,keepaspectratio]{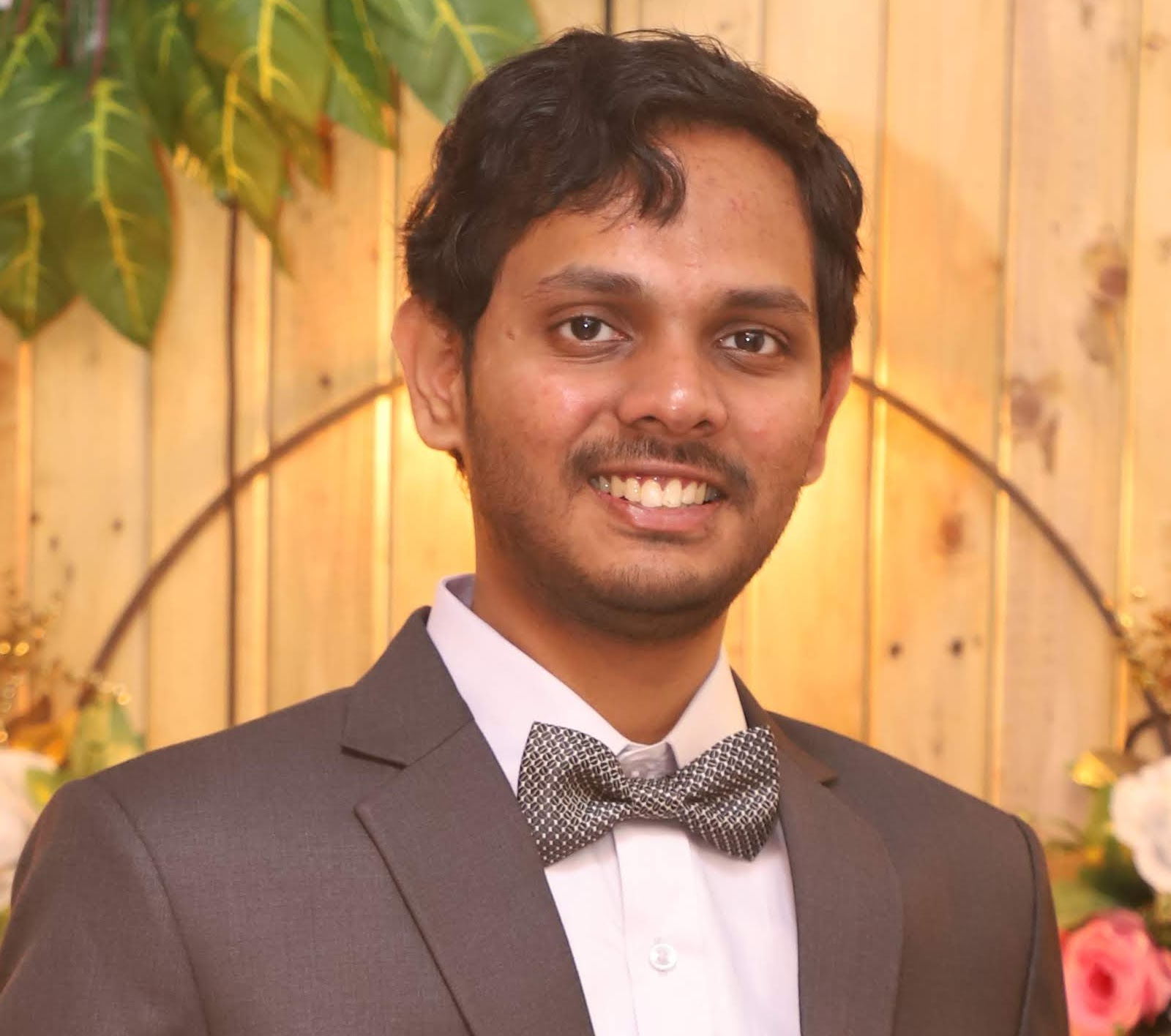}}]{Koustubh Phalak} is a Ph.D. student at the Department of Computer Science and Engineering in Pennsylvania State University. He completed his Bachelors in Electrical and Electronics Engineering from Birla Institute of Technology and Science, Pilani in 2020. He works in the field of emerging technologies, specially quantum computing.
\end{IEEEbiography}
\vskip 0pt plus -1fil

\begin{IEEEbiography}[{\includegraphics[width=1in,height=1.25in,clip,keepaspectratio]{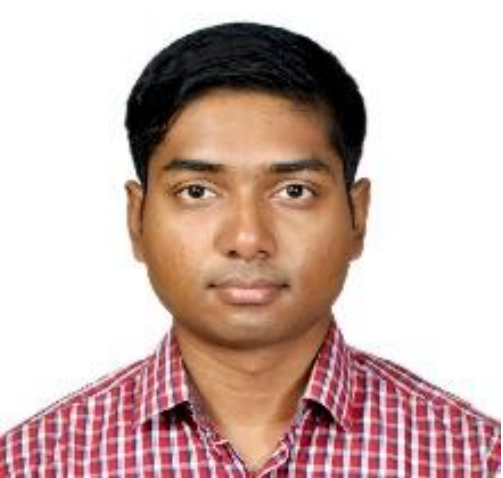}}]{Abdullah Ash- Saki (S'12)} is a doctoral student at the Department of Electrical Engineering in Pennsylvania State University. He received his Bachelors from Bangladesh University of Engineering and Technology (BUET) in 2014. He was the recipient of Dr. Nirmal K. Bose Dissertation Excellence Award from Dept. of Electrical Engineering at Penn State, ECE Best Paper Award in 2020 ASEE Annual Conference and, Dr. Richard Newton Young Fellow award in 55th Design Automation Conference (DAC). He works on quantum computing. 
\end{IEEEbiography}
\vskip 0pt plus -1fil

\begin{IEEEbiography}[{\includegraphics[width=1in,height=1.25in,clip,keepaspectratio]{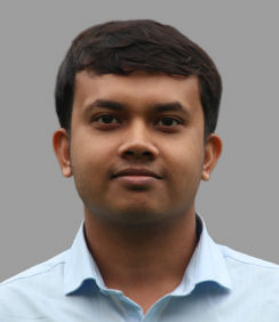}}]{Mahabubul Alam} received his B.Sc. degree in Electrical and Electronic Engineering from Bangladesh University of Engineering and Technology (BUET), in 2015, and currently pursuing his PhD degree in Electrical Engineering at Pennsylvania State University. Before joining the graduate school, he worked as an ASIC Physical Design Engineer at PrimeSilicon Technologies for over a year. He spent the summer of 2018 and 2020 as an intern at Qualcomm and Intel, respectively. He received the best paper award (BPA) at ASEE (2020), BPA nomination at ISQED (2020), and secured second place at the student research competition at ICCAD (SRC@ICCAD-2020). His current research interests include quantum computing, machine learning, design automation, and hardware security.
\end{IEEEbiography}
\vskip 0pt plus -1fil

\begin{IEEEbiography}[{\includegraphics[width=1in,height=1.25in,clip,keepaspectratio]{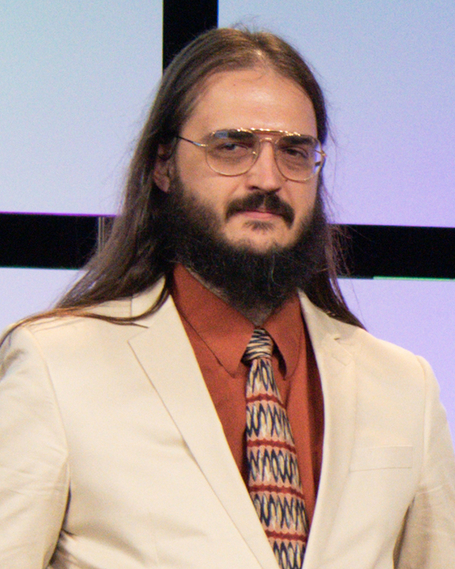}}]{Rasit Onur Topaloglu (M'05–SM'13)} obtained his B.S. in EE from Bogazici University and Ph.D. in Computer Science and Engineering from University of California at San Diego. He has worked for companies such as Qualcomm, AMD, GLOBALFOUNDRIES and is currently with IBM. He works on next-generation computer technology and design currently as a Senior Hardware Developer. He was partially involved with qubit characterization laboratory work at IBM Research. He has over sixty peer-reviewed publications and over sixty issued US patents, more than a third of which are on quantum technologies. He has chaired the IEEE/ACM DAC Workshop on Design Automation for Quantum (DAQ). As of 2021, he is working on a Quantum Computing book. He serves on IEEE/ACM Design Automation Conference (DAC), IEEE/ACM International Conference on Computer-Aided Design (ICCAD), and IEEE International Symposium on Quality Electronic Design (ISQED) Technical Program Committees that cover quantum topics. He serves as the Chair of IEEE Mid-Hudson and the Secretary of ACM Poughkeepsie. He is an IEEE/ACM DAC Outstanding Innovator and an IBM Master Inventor.
\end{IEEEbiography}
\vskip 0pt plus -1fil

\begin{IEEEbiography}[{\includegraphics[width=1in,height=1.25in,clip,keepaspectratio]{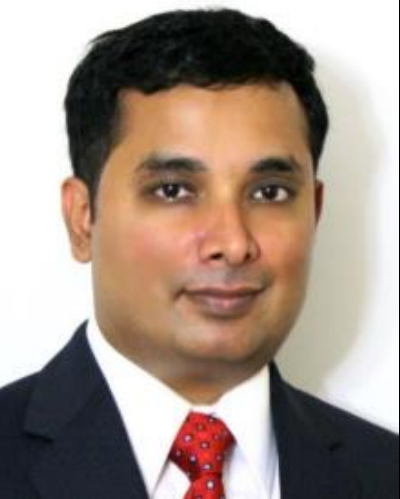}}]{Swaroop Ghosh (SM'13)} received the B.E. (Hons.) from IIT, Roorkee and the Ph.D. degree from Purdue University.
He is an Associate Professor at Pennsylvania State University.
His research interests include quantum computing, emerging memory technologies and hardware security.
			
Dr. Ghosh served as Associate Editor of the IEEE Transactions On Circuits and Systems I and IEEE Transactions On Computer-Aided Design and as Senior Editorial Board member of IEEE Journal of Emerging
Topics on Circuits and Systems (JETCAS). He served as Guest Editor of the IEEE JETCAS and IEEE Transactions On VLSI Systems. He has also served in the technical program committees of more than 25 ACM/IEEE conferences. He served as General Chair, Conference Chair and Program Chair of of ISQED and DAC Ph.D. Forum and track (co)-Chair in DAC, CICC, ISLPED, GLSVLSI, VLSID and ISQED.

Dr. Ghosh is a recipient of Intel Technology and Manufacturing Group Excellence Award, Intel Divisional Award, two Intel Departmental Awards, USF Outstanding Research Achievement Award, College of Engineering Outstanding Research Achievement Award, DARPA Young Faculty Award (YFA), ACM SIGDA Outstanding New Faculty Award, YFA 	Director’s Fellowship, Monkowsky Career Development Award, Lutron Spira Teaching Excellence Award, Dean's Certificate of Excellence and Best Paper Award in American Society of Engineering Education (ASEE). He is a Senior member of the IEEE and the National Academy of Inventors (NAI), Associate member of Sigma Xi and Distinguished Speaker of the Association for Computing Machinery (ACM).

\end{IEEEbiography}

\end{document}